\documentclass[aps,11pt,twoside, nofootinbib]{revtex4}
\usepackage{amsthm}
\usepackage{amsmath,latexsym,amssymb,verbatim,enumerate,graphicx}
\usepackage{hyperref}
\usepackage{color}

\newtheorem{thm}{Theorem}
\newtheorem*{thm*}{Theorem}

\newtheorem*{prop*}{Proposition}

\newtheorem*{lem*}{Lemma}

\newtheorem*{fact*}{Fact}

\newtheorem*{cor*}{Corollary}
\newtheorem{conjecture}[thm]{Conjecture}

\newtheorem*{rep@theorem}{\rep@title}
\newcommand{\newreptheorem}[2]{%
\newenvironment{rep#1}[1]{%
 \def\rep@title{#2 \ref{##1} (restatement)}%
 \begin{rep@theorem}}%
 {\end{rep@theorem}}}
\makeatother

\newreptheorem{thm}{Theorem}
\newreptheorem{lem}{Lemma}
\newreptheorem{cor}{Corollary}

\usepackage{prettyref}
\newcommand{\savehyperref}[2]{\texorpdfstring{\hyperref[#1]{#2}}{#2}}

\newrefformat{eq}{\savehyperref{#1}{\textup{(\ref*{#1})}}}
\newrefformat{lem}{\savehyperref{#1}{Lemma~\ref*{#1}}}
\newrefformat{def}{\savehyperref{#1}{Definition~\ref*{#1}}}
\newrefformat{thm}{\savehyperref{#1}{Theorem~\ref*{#1}}}
\newrefformat{cor}{\savehyperref{#1}{Corollary~\ref*{#1}}}
\newrefformat{cha}{\savehyperref{#1}{Chapter~\ref*{#1}}}
\newrefformat{sec}{\savehyperref{#1}{Section~\ref*{#1}}}
\newrefformat{app}{\savehyperref{#1}{Appendix~\ref*{#1}}}
\newrefformat{tab}{\savehyperref{#1}{Table~\ref*{#1}}}
\newrefformat{fig}{\savehyperref{#1}{Figure~\ref*{#1}}}
\newrefformat{hyp}{\savehyperref{#1}{Hypothesis~\ref*{#1}}}
\newrefformat{alg}{\savehyperref{#1}{Algorithm~\ref*{#1}}}
\newrefformat{rem}{\savehyperref{#1}{Remark~\ref*{#1}}}
\newrefformat{item}{\savehyperref{#1}{Item~\ref*{#1}}}
\newrefformat{step}{\savehyperref{#1}{step~\ref*{#1}}}
\newrefformat{conj}{\savehyperref{#1}{Conjecture~\ref*{#1}}}
\newrefformat{fact}{\savehyperref{#1}{Fact~\ref*{#1}}}
\newrefformat{prop}{\savehyperref{#1}{Proposition~\ref*{#1}}}
\newrefformat{prob}{\savehyperref{#1}{Problem~\ref*{#1}}}
\newrefformat{claim}{\savehyperref{#1}{Claim~\ref*{#1}}}
\newrefformat{relax}{\savehyperref{#1}{Relaxation~\ref*{#1}}}
\newrefformat{red}{\savehyperref{#1}{Reduction~\ref*{#1}}}
\newrefformat{part}{\savehyperref{#1}{Part~\ref*{#1}}}

\newcommand{\MYstore}[2]{%
  \global\expandafter \def \csname MYMEMORY #1 \endcsname{#2}%
}

\newcommand{\MYload}[1]{%
  \csname MYMEMORY #1 \endcsname%
}

\newcommand{\MYnewlabel}[1]{%
  \newcommand\MYcurrentlabel{#1}%
  \MYoldlabel{#1}%
}

\newcommand{\MYdummylabel}[1]{}

\newcommand{\torestate}[1]{%
  \let\MYoldlabel\label%
  \let\label\MYnewlabel%
  #1%
  \MYstore{\MYcurrentlabel}{#1}%
  \let\label\MYoldlabel%
}

\newcommand{\restatethm}[1]{%
  \let\MYoldlabel\label
  \let\label\MYdummylabel
  \begin{repthm}{#1}
    \MYload{#1}
  \end{repthm}
  \let\label\MYoldlabel
}

\newcommand{\restatelem}[1]{%
  \let\MYoldlabel\label
  \let\label\MYdummylabel
  \begin{replem}{#1}
    \MYload{#1}
  \end{replem}
  \let\label\MYoldlabel
}

\newcommand{\restatelemalt}[1]{%
  \let\MYoldlabel\label
  \let\label\MYdummylabel
  \begin{lem*}[Restatement of \prettyref{#1}]
    \MYload{#1}
  \end{lem*}
  \let\label\MYoldlabel
}

\newcommand{\restatecor}[1]{%
  \let\MYoldlabel\label
  \let\label\MYdummylabel
  \begin{repcor}{#1}
    \MYload{#1}
  \end{repcor}
  \let\label\MYoldlabel
}

\newcommand{\restateprop}[1]{%
  \let\MYoldlabel\label
  \let\label\MYdummylabel
  \begin{prop*}[Restatement of \prettyref{#1}]
    \MYload{#1}
  \end{prop*}
  \let\label\MYoldlabel
}

\newcommand{\restatefact}[1]{%
  \let\MYoldlabel\label
  \let\label\MYdummylabel
  \begin{fact*}[Restatement of \prettyref{#1}]
    \MYload{#1}
  \end{fact*}
  \let\label\MYoldlabel
}

\newcommand{\restate}[1]{%
  \let\MYoldlabel\label
  \let\label\MYdummylabel
  \MYload{#1}
  \let\label\MYoldlabel
}

\newtheorem{Def}{Definition}
\newtheorem{Thm}[Def]{Theorem}
\newtheorem{Lem}[Def]{Lemma} 
\newtheorem{Ex}[Def]{Example}
\newtheorem{Cor}[Def]{Corollary}
\newtheorem{Rem}[Def]{Remark}
\newcommand{\bdf}{\begin{Def}}
\newcommand{\edf}{\end{Def}}
\newcommand{\bex}{\begin{Ex}}
\newcommand{\eex}{\end{Ex}}
\newcommand{\bthm}{\begin{Thm}}
\newcommand{\ethm}{\end{Thm}}
\newcommand{\blm}{\begin{Lem}}
\newcommand{\elm}{\end{Lem}}
\newcommand{\bcor}{\begin{Cor}}
\newcommand{\ecor}{\end{Cor}}
\newcommand{\brem}{\begin{Rem}}
\newcommand{\erem}{\end{Rem}}
\renewcommand{\>}{\rangle}
\newcommand{\ot}{\otimes}

\newcommand*{\cD}{\mathcal{D}}
\newcommand*{\cE}{\mathcal{E}}

\newcommand*{\cH}{\mathcal{H}}

\newcommand*{\cO}{\mathcal{O}}

\newcommand*{\cV}{\mathcal{V}}

\newcommand*{\veps}{\varepsilon}
\newcommand*{\trho}{{\tilde \rho}}

\newcommand*{\tO}{{\tilde O}}

\newcommand*{\tE}{{\tilde E}}
\newcommand*{\tpi}{{\tilde \pi}}

\newcommand*{\Th}{\Theta}
\newcommand*{\tLambda}{{\tilde \Lambda}}

\newcommand*{\tr}{\mathrm{Tr}}

\newcommand*{\argmax}{\mathop{\rm arg~max}\limits}
\newcommand*{\id}{\rm id}

\def\idty{{\leavevmode\rm 1\mkern -5.4mu I}}

\usepackage{color}
\definecolor{myred}{rgb}{1,0,0}

\definecolor{myblue}{rgb}{0,0,0.8}

\definecolor{myyellow}{rgb}{0.9,0.8,0}

\definecolor{mygreen}{rgb}{0,0.6,0}

\definecolor{myorange}{rgb}{0.6,0.6,0}

\definecolor{mycerul}{rgb}{0,0.6,1}

\begin{document}

\title{{\Large Quantum Approximate Markov Chains are Thermal}}

\author{Kohtaro Kato}
\affiliation{Department of Physics, Graduate School of Science, The University of Tokyo, Tokyo, Japan}
\affiliation{Institute for Quantum Information and Matter  \\ California Institute of Technology, Pasadena, CA 91125, USA}

\author{Fernando G.S.L. Brand\~ao}
\affiliation{Institute for Quantum Information and Matter  \\ California Institute of Technology, Pasadena, CA 91125, USA}

\begin{abstract}
We prove that any one-dimensional (1D) quantum state with small quantum conditional mutual information in all certain tripartite
splits of the system, which we call a {\it quantum approximate Markov chain}, can be well-approximated by a Gibbs state of a short-range quantum Hamiltonian. 
Conversely, we also derive an upper bound on the (quantum) conditional mutual information of Gibbs states of 1D 
short-range quantum Hamiltonians. We show that the conditional mutual
information between two regions $A$ and $C$ conditioned on the middle region $B$ decays
exponentially with the square root of the length of $B$. 

These two results constitute a variant of the Hammersley-Clifford theorem (which characterizes Markov networks, i.e. probability distributions
which have vanishing conditional mutual information, as Gibbs states of classical short-range Hamiltonians) for 1D quantum systems. The result can be seen as a strengthening - for 1D systems - of the mutual information area law for thermal states. It directly implies an efficient preparation of any 1D Gibbs state at finite temperature by a constant-depth quantum circuit.
\end{abstract}

\maketitle

\parskip .75ex


\section{Introduction}

A sequence of discrete random variables $X_1, \ldots, X_n$ forms a Markov chain if $X_{i+1}$ is uncorrelated from $X_{1}, \ldots, X_{i-1}$ conditioned on the value of $X_i$. Markov chains are a central concept in probability theory, statistics and beyond. In this paper we consider a combination of two natural generalizations of the concept of a Markov chain. 

In the first we only require {\it approximate} independence from previous random variables, i.e. $X_{i+1}$ should only be \textit{almost} independent from $X_{1}, \ldots, X_{i-1}$ conditioned on $X_i$. One way to make this notion quantitative is to use the conditional mutual information, defined for every three random variables $X, Y, Z$ drawn from the distribution $p(X, Y, Z)$ as 
\begin{equation*}
I(X : Z|Y)_p := H(XY)_p + H(YZ)_p - H(XYZ)_p - H(Y)_p\,,
\end{equation*}
where $H(X)_p:=-\sum_{x_i\in X} p_X(x_i)\log p_X(x_i)$ is the Shannon entropy of the marginal distribution on $X$~\footnote{In the following we always use 2 as the base of log.}. 
In terms of the conditional mutual information, $X_1, \ldots, X_n$ is a Markov chain if
\begin{equation*}\label{standardmc}
I(X_1\ldots X_{i-1} : X_{i+1} \ldots X_n | X_i)_p=0 \hspace{0.2 cm}  \forall i \in [1, n]\,.
\end{equation*}
The conditional mutual information can also be written as 
\begin{equation}  \label{conditioned}
I(X : Z | Y)_{p} = \mathbb{E}_{Y \sim p(Y)} I(X : Z)_{p_{y}}\,, 
\end{equation}
where $I(X:Z)_{p_y}$ is the mutual information: 
\begin{equation}\label{eq:defmi}
I(X:Z)_{p_y}:=H(X)_{p_y}+H(Z)_{p_y}-H(XZ)_{p_y}\,,
\end{equation}
with $p_{y}(X, Z)$ which is the conditional distribution of $X$ and $Z$ for given $Y = y$. Thus if $I(X : Z|Y)_p$ is small, $X$ and $Z$ are almost uncorrelated conditioned on $Y$. 
For $\veps>0$, we say $X_1, \ldots, X_n$ is a {\it $\varepsilon$-approximate Markov chain} if 
\begin{equation*}
I(X_1\ldots X_{i-1} : X_{i+1} \ldots X_n | X_i)_p \leq \varepsilon \hspace{0.2 cm}  \forall i \in [1, n]\,.
\end{equation*}

In the second, instead of considering random variables, we consider a $n$-partite \textit{quantum} state given by a density matrix $\rho_{A_1 \ldots A_n} \in {\cal D}({\cal H}_{A_1} \otimes \ldots \otimes {\cal H}_{A_n})$ \footnote{${\cal D}({\cal H}_{A_1} \otimes \ldots \otimes {\cal H}_{A_n})$ is the set of density matrices over the finite-dimensional Hilbert space ${\cal H}_{A_1} \otimes \ldots \otimes {\cal H}_{A_n}$.}. The quantum conditional mutual information of a tripartite state $\rho_{ABC}$ is defined as 
\begin{equation*}
I(A : C|B)_{\rho} := S(AB)_{\rho} + S(BC)_{\rho} - S(ABC)_{\rho} - S(B)_{\rho}\,, 
\end{equation*}
where $S(X)_{\rho} := - \tr(\rho_X \log \rho_X)$ is the von Neumann entropy of the reduced state on subsystem $X$. Quantum states satisfying $I(A:C|B)_\rho = 0$ are analogues of Markov chains of three random variables. 
As in the classical case~\eqref{standardmc}, a multipartite quantum state $\rho_{A_1, \ldots, A_n}$ is a quantum Markov chain if 
\begin{equation*}
I(A_1\ldots A_{i-1} : A_{i+1} \ldots A_n | A_i)_{\rho} = 0  \hspace{0.2 cm} \forall i \in [1,n]. 
\end{equation*}

In this paper, we are interested in {\it quantum approximate Markov chains},  a combination of both generalizations. Such concept is already non-trivial for tripartite quantum states $\rho_{ABC}$. We can say $\rho_{ABC}$ forms a quantum $\varepsilon$-approximate Markov chain if 
\begin{equation*} 
I(A:C|B)_{\rho} \leq \varepsilon.
\end{equation*}
However there is no quantum analogue of Eq.\;\eqref{conditioned} (see Ref.\;\cite{ibinson2008robustness}) and therefore it is unclear if the definition in the above has a nontrivial meaning. A recent result in quantum information theory reveals its meaning~\cite{Fawzi2015}. It shows that
\begin{equation} \label{fawzirenner}
I(A:C|B)_{\rho}  \geq \min_{ \Delta : B \rightarrow BC} -2 \log F(\rho_{ABC}, \Delta_{B \rightarrow BC}(\rho_{AB})),
\end{equation}
where the minimum is over all completely-positive and trace-preserving (CPTP) map $\Delta_{B \to BC}$ mapping $\cD({\cal H}_B)$ to $\cD({\cal H}_B \otimes {\cal H}_C)$, and $F(\rho, \sigma) := \tr((\sigma^{1/2} \rho \sigma^{1/2})^{1/2})$ is the fidelity. Thus if the conditional mutual information is small, $A$ is only correlated to $C$ through $B$ up to a small error, in the sense that $C$ can be approximately recovered given the information contained in $B$ only (see Refs.\;\cite{PhysRevLett.100.230501,2016arXiv160906994B,2017arXiv170302903S}). More generally, we say $\rho_{A_1, \ldots, A_n}$ is a quantum $\varepsilon$-approximate Markov chain if 
\begin{equation*}
I(A_1, \ldots, A_{i-1} : A_{i+1}, \ldots, A_n | A_i)_{\rho} \leq \varepsilon  \hspace{0.2 cm} \forall i \in [1, n]\,. 
\end{equation*}

\subsection{The Hammersley-Clifford theorem}

In this paper we will be interested in finding a structural characterization of quantum approximate Markov chains. Our motivation is a powerful result in statistics called the Hammersley-Clifford Theorem~\cite{HCthm71}. It states that Markov chains (and more generally Markov networks \footnote{A Markov network is a generalization of a Markov chain given by random variables $X_{1}, \ldots, X_{n}$ defined on the vertices $1, \ldots, n\in V$ of a graph $G=(V,E)$, such that $X_{i}$ is uncorrelated from all other random variables conditioned on the random variables $\{ X_{j}  \}_{(i,j)\in E }$ associated to neighboring vertices.}), in which all elements of the distribution are non-zero, are equivalent to the set of Gibbs (thermal) states of nearest-neighbor Hamiltonians on a 1D open spin chain~\footnote{For Markov networks, in turn, the Hamiltonian is a sum of local functions of variables on all cliques of the graph.}: 
\begin{equation*}
p(X_1 = x_1, \ldots, X_n = x_n) = \frac{1}{Z} \exp \left( -\sum_i h_{i}(x_{i}, x_{i+1})   \right),
\end{equation*}
for functions $h_i : \mathbb{R}^{2} \rightarrow \mathbb{R}$, where
\begin{equation*}
Z := \sum_{x_1, \ldots, x_n}\exp \left( -\sum_i h_{i}(x_{i}, x_{i+1})   \right)
\end{equation*}
is the partition function. Here, the ``temperature'' is included in the interaction terms. 

In Refs.~\cite{Leifer20081899, 2007quant.ph..1029Z}, the Hammersley-Clifford theorem was generalized to quantum Markov chains (and Markov networks): A full-rank quantum state $\rho_{A_1 \ldots A_n}$ is a quantum Markov chain if, and only if, it can be written as
\begin{equation*}
\rho_{A_1...A_n} = \frac{1}{Z}\exp\left(-\sum_i h_{i, i+1}\right),
\end{equation*}
where $Z=\tr(\exp(-\sum_i h_{i,  i+1}))$ and each $h_{i, i+1}$ only acts on subsystems $A_iA_{i+1}$, such that $[h_{i, i+1}, h_{j, j+1}] = 0$ for all $i, j$. Therefore we have a characterization of full-rank quantum Markov chains as Gibbs states of 1D commuting short-range quantum Hamiltonians \footnote{In Ref.~\cite{2007quant.ph..1029Z} a more general result was shown for quantum Markov networks. In contrast to the classical case, positive (i.e. full-rank) quantum Markov networks are only equivalent to Gibbs states of commuting Hamiltonians with terms on the cliques of the graph if the graph is triangle free.}. 
Conversely, this result also clarifies that correlations in Gibbs states of 1D commuting short-range Hamiltonians are always mediated through interactions between neighboring regions.

The characterization above only involves exact quantum Markov chains and a special set of short-range Hamiltonians. 
A natural question is whether there is a similar relation between quantum approximate Markov chains and more general quantum Gibbs states.  
The main result of this paper answers the question in the affirmative: we prove that quantum approximate Markov chains are equivalent to Gibbs states of 1D short-range quantum Hamiltonians, which we also call {\it local Gibbs states} in short.

\noindent \textbf{Notation:} 
In the following, we consider a quantum spin system $\Lambda$ on a graph $G=(V,E)$, where $V=\{1,...,n\}$ and $E=\{(i,i+1)\}_{i=1}^{n-1}$ for $n\in\mathbb Z_{>0}$, i.e., a 1D open spin chain. Sometimes we also consider a closed chain by adding additional edge $(n,1)$. 
The Hilbert space of a local subsystem $\cH_i$ corresponding to spin $i$ is associated to each $i\in V$, which has finite dimension $d<\infty$.  
For a subsystem specified by subregion $X\subset V$ (we abuse the notation by using same $X$ to denote both a subsystem and a subregion), we denote $\cH_X:=\bigotimes_{i\in X}\cH_{i}$. The logarithm of $\dim\cH_{X}$ is simply denoted by $|X|$. 
We denote the operator norm of an operator $O$ by $\|O\|$, and the trace norm of $O$ by $\|O\|_1$. 
We say that the support of $O$ is $X$ if an operator $O$ on $\Lambda$ can be written as 
\begin{equation*}
O=O_X\ot\idty_{X^c}\,,
\end{equation*}
i.e., the tensor product of some operator $O_X$ on  region $X$ and the identity operator $\idty_{X^c}$ acting on the complement of $X$ (which we denoted by $X^c$). We will denote the support of an operator $O$ by $\text{supp}(O)$, unless explicitly mentioned. 

A short-range Hamiltonian $H$ is a bounded Hermitian operator on $\cH_\Lambda$ which can be decomposed into
\begin{equation*}
H=\sum_i h_i\,,
\end{equation*}
where each $\|h_i\|$ is bounded by a constant and $\text{supp}(h_i)$ only contains spins within graph distance $r<\infty$ from the spin $i$. 
We also consider a restricted Hamiltonian $H_X$ on a region $X$, defined as 
\begin{equation*}
H_X=\sum_{\text{supp}(h_i)\subset X}h_i\,,
\end{equation*}
i.e., the sum of interactions acting on spins sitting {\it inside} of $X$. The Gibbs state $\rho^{H_X}$ of the Hamiltonian $H_X$ at (inverse) temperature $\beta>0$ is defined as
\begin{equation}\label{def:gibbs}
\rho^{H_X}=\frac{e^{-\beta H_X}}{Z_X}\,,
\end{equation}
where $Z_X=\tr[e^{-\beta H_X}]$. Note that we will omit $\beta$ in the notations and simply denote $\rho^{H_X}$ and $Z_X$, while they depend on $\beta$.
 The reduced state of the Gibbs state on a subregion $Y\subset X$ is denoted by $\rho^{H_X}_Y$. 
 
Throughout the paper, we often consider a disjoint tripartition $ABC$ of $\Lambda$. We say
$B$ shields $A$ from $C$ if $A$ and $C$ are indirectly connected through $B$.  We denote by $d(A, C)$ the graph distance between $A$ and $C$. $d(A, C)=|B|$ if $B$ shields $A$ from $C$ and is connected. 

\section{Results}
In this section we present the main results of this paper. 
For two quantum states $\rho, \sigma$, we denote their (quantum) relative entropy as
\begin{equation*}
S(\rho \Vert \sigma) := \tr(\rho(\log \rho - \log \sigma)).
\end{equation*}
if $\text{supp}(\rho)\subset \text{supp}(\sigma)$, and $S(\rho \Vert \sigma) :=+\infty$ otherwise (Here \lq\lq{}$\text{supp}(\rho)$\rq\rq{} means the subspace spanned by eigenvectors of $\rho$ with nonzero eigenvalues). 

\subsection{Approximation of quantum approximate Markov states by local Gibbs states}

Let us divide 1D spin chain $\Lambda$ into $m$ connected regions $A_1A_2...A_m$. We denote the coarse-grained 1D spin chain $A_1A_2...A_m$ by $A$. 
Our first result is the following theorem (see Section~\ref{sec:proof2} for the proof): 
\bthm\label{mthm1}
Let $\rho_{A_1 \ldots A_m}$ be a quantum $\varepsilon$-approximate Markov chain on a 1D open chain for $\veps>0$. Then there exists a short-range Hamiltonian $H = \sum_{i=1}^{m-1} h_{A_i, A_{i+1}}$ with $ \text{supp}(h_{A_i, A_{i+1}})=A_iA_{i+1}$, such that 
\begin{equation}\label{eq:thm1}
S \left( \rho \left \Vert  \frac{e^{-H}}{Z} \right.  \right)\leq \varepsilon m\,,
\end{equation} 
where $Z=\tr e^{-H}$. 
\ethm
From the relation
\begin{equation*}
S(\rho \Vert \sigma) \geq - 2 \log F(\rho, \sigma),
\end{equation*}
we find that the state $\rho$ is also close in fidelity to a local Gibbs state. Note that Theorem 1 is not restricted to full-rank states. 

It is natural to expect that there exists a similar bound for 1D closed chains. 
In this paper, we say a state $\rho_{A_1...A_m}$ is a quantum $\veps$-approximate Markov chain on a 1D {\it closed} chain if
\begin{equation}\label{mcring}
I(A_i: A\backslash A_{i-1}A_iA_{i+1} | A_{i-1}A_{i+1})_\rho \leq \varepsilon, \hspace{0.2 cm} \forall i \in [1,m]. 
\end{equation}
in analogy with a quantum Markov network~\footnote{Any quantum $\veps$-approximate Markov chain $\rho_{A_1...A_m}$ satisfies
$I(A_i: A\backslash A_{i-1}A_iA_{i+1} | A_{i-1}A_{i+1})_\rho \leq 2\varepsilon$ for all $i \in [3,m-2]$. }. 
Here, we imposed the periodic boundary condition on labels, e.g., $m+1\equiv1$. 
In this situation, the proof of Theorem~\ref{mthm1} does not work straightforwardly. 
To solve this difficulty,  we consider two sufficient conditions as assumptions which show the closeness to local Gibbs states respectively. 

The first assumption is that the exsitence of the finite correlation length in terms of the quantum mutual information, which is obtained by replacing the Shannon entropy by the von Neumann entropy in Eq.~\eqref{eq:defmi}. 
The second assumption is called the uniform Markov property~\cite{2016arXiv160907877B} requiring that the reduced states of the state are also approximately Markov chains.  
Note that the second assumption may be derived from the definition of the quantum approximate Markov chains while we have not had any result on it. 
\bthm\label{thm2}
Let $\rho_{A_1 \ldots A_m}$ be a quantum $\varepsilon$-approximate Markov chain on a 1D closed chain for $\veps>0$. 
\begin{itemize}
\item[$(i)$]  
Assume that $\rho_{A_1\ldots A_m}$ also satisfies 
\begin{equation*}
I(A_i:A\backslash A_{i-1}A_iA_{i+1})_\rho\leq\veps \hspace{0.2cm}\forall i\in[1,m]\,.
\end{equation*}
Then there exists a short-range Hamiltonian $H = \sum_{i} h_{A_{i-1},A_i, A_{i+1}}$, with $\text{supp}(h_{A_{i-1},A_i, A_{i+1}})=A_{i-1}A_i A_{i+1}$, such that 
\begin{equation*}
S \left( \rho \left \Vert  \frac{e^{-H}}{Z} \right.  \right)\leq \varepsilon m. 
\end{equation*}

\item[$(ii)$]
Assume that for any $i\in[1,m]$, $\tr_{A_i}(\rho_{A_1\ldots A_m})$ is a quantum $\veps$-approximate Markov chain  for the 1D open chain $A_{i+1}A_{i+2}...A_{i-1}$ (we used $m+1\equiv 1$).  
Then there exists a short-range Hamiltonian $H = \sum_{i} h_{A_{i-1},A_i, A_{i+1}}$, with $\text{supp}(h_{A_{i-1},A_i, A_{i+1}})=A_{i-1}A_i A_{i+1}$, such that 
\begin{equation*}
S \left( \rho \left \Vert  \frac{e^{-H}}{Z} \right.  \right)\leq \varepsilon m. 
\end{equation*}
\end{itemize}
\ethm

The approximate Markov property appears in analysis of gapped ground states of many-body systems (see e.g., Refs.~\cite{PhysRevB.87.155120, Flammia2017limitsstorageof}). 
In these cases, the additional assumptions seem to be satisfied for certain choice of subsystems. 
Interestingly, there exists a class of states which are locally quantum approximate Markov chains but globally not. 
A simple example is the $n$-qubit GHZ state
\begin{equation*}
|GHZ_n\>=\frac{1}{\sqrt{2}}\left(|0000...0\>+|1111...1\>\right)\,
\end{equation*}
on a spin chain (either open or closed). When we trace out one qubit from the chain, the reduced state exactly become a quantum Markov chain. 
However,  this state has $I(A:C|B)=1$ for any tripartition $ABC$ of the whole system where $B$ shields $A$ from $C$.  
Therefore, it is not a quantum Markov chain globally. 
A similar situation arises when considering a ring-like regions in systems with topological order~\cite{PhysRevLett.96.110405}. 
We show that for these cases, the value of the conditional mutual information for the whole system approximately 
represents the distance from the set of local Gibbs states. We will discuss an application of this result for analysis of entanglement spectrum in 2D topologically ordered phases in a complementary work~\cite{upcoming17}. 
\bthm\label{cmirelative}
Consider a 1D spin chain $X=X_1X_2...X_m$ with the size $N=|X_1...X_m|$. Let $\rho_{X_1...X_m}$ be a state such that  the reduced state obtained by tracing out $X_i$ is a quantum $\varepsilon$-approximate Markov chain  for all $i\in[1, m]$.   
Define the set of Gibbs states of short-range Hamiltonians with interaction strength $K$ as 
\begin{equation*}
E^K_{nn}:=\left\{e^{-H}\left|\; H=\sum_ih_{X_iX_{i+1}},\; \tr(e^{-H})=1,\,\;\|h_{X_iX_{i+1}}\|\leq K\right.\right\}\,.
\end{equation*}
Note that here we include the normalization factor in the Hamiltonians. 
Then, for $K=\Theta(N)$ and sufficiently small $\veps>0$, there exists a constant $c>0$ such that  for any tripartition $ABC$ of the whole system such that $B$ shields $A$ from $C$, it holds that
\begin{equation}\label{eq:thm4}
\min_{\mu\in E^K}S\left(\rho_{X}\left\| \mu\right.\right)=I(A:C|B)_\rho +\epsilon(N,\delta)
\end{equation}
and
\begin{equation*}
\left|\epsilon(N,\delta)\right|\leq cN^{\frac{5}{2}}\delta^{\frac{1}{16}}\,,
\end{equation*}   
where $\delta=8\sqrt{\veps}+2^{-N}$. 
\ethm
Here we used $X_i$ to label subsystems instead of $A_i$ to avoid confusion with $A\subset X$. 

\subsection{Quantum approximate Markov property of 1D Gibbs states}
Our second main result is a kind of converse to Theorem~\ref{mthm1} and Theorem~\ref{thm2}.
(see Section~\ref{sec:proof} for the proof): 
\bthm\label{m.mthm2}
Let $H = \sum_i h_i$ be a short-range 1D Hamiltonian with $\Vert h_i \Vert \leq 1$ and $l_0, c, c'>0$ be universal constants. For an inverse temperature $\beta > 0$ and any partition $ABC$ with $d(A, C)\geq l_0$,  there exists a CPTP-map $\Lambda_{B\to BC} : \cD({\cal H}_B) \rightarrow  \cD({\cal H}_B  \otimes  {\cal H}_C)$ such that
\begin{equation}\label{mthm2}
\left \Vert  \rho^{H_{ABC}}  -\Lambda_{B\to BC}(\rho^{H_{ABC}}_{AB}) \right \Vert_1 \leq e^{- q(\beta) \sqrt{d(A,C)}}\,,
\end{equation}
where $q(\beta) =  ce^{- c' \beta }$ if the correlation length of $\rho^{H_{ABC}}$ is $\xi=e^{\cO(\beta)}$. 
\ethm

The theorem above states that if we choose the region $B$ sufficiently large, the Gibbs state can be approximately recovered from the partial trace over $C$ by performing a recovery map on $B$. In turn, the statement implies the corresponding conditional mutual information decays similarly:
\bcor~\label{cor1}
Under the setting of Theorem~\ref{m.mthm2}, 
\begin{equation*}
I(A:C|B)_{\rho^{H_{ABC}}}\leq 6\left(d(A,C)+\frac{8(1+\frac{q(\beta) }{2}\sqrt{d(A,C)})}{q(\beta)^2}\right) e^{- \frac{q(\beta)}{2} \sqrt{d(A,C)}}.
\end{equation*}
\ecor
Therefore, $I(A:C|B)_\rho\leq \cO(e^{-\Th(\sqrt{d(A,C)})})$ and thus any 1D local Gibbs state is a quantum approximate Markov chain with small $\veps$ after certain coarse-graining. Conversely, Theorem \ref{mthm1}  shows that any quantum approximate Markov chain can be well-approximated by some 1D local Gibbs state. Therefore the combination of the two results can be regarded as a variant of the Hammersley-Clifford theorem for quantum approximate Markov chains. Below we discuss two implications of our results:

\subsection{Saturation rate of area law for 1D Gibbs states} \label{sec:satuarea}

A Gibbs state of a short-range Hamiltonian obeys an area law in terms of the mutual information. 
For instance, for any Gibbs state $\rho^H$ of a short-range Hamiltonian on a lattice, it holds that~\cite{PhysRevLett.100.070502}
\begin{equation}\label{areami}
I(A:A^c)_{\rho^H}\leq \beta J|\partial A|\,,
\end{equation}
where $J>0$ is a constant only depending on the locality of $H$ and the norm of the local interactions. The upper bound is a constant when the system is 1D. 

The area law represented by Eq.~\eqref{areami} implies a decay of the conditional mutual information. 
Consider a 1D spin chain system and let $A$ be a simply connected region. We divide $A^c$ into $\{b_i\}_i$ so that $b_1$ shields $b_2$ from $A$, $b_2$ shields $b_3$ from $Ab_1$ and so on.  We set $B_l=b_1b_2\ldots b_l$. By the monotonicity of the mutual information under the partial trace, we have 
\begin{equation*}
I(A:B_l)_{\rho^H}\leq I(A:B_{l+1})_{\rho^H}\leq I(A:A^c)_{\rho^H}\leq \beta J|\partial A|\,.
\end{equation*}
Since the upper bound is independent of $l$, $I(A:b_{l+1}|B_{l})_{\rho^H}=I(A:B_{l+1})_{\rho^H}-I(A:B_l)_{\rho^H}$ eventually vanishes when $l$ grows. 
Corollary \ref{cor1} goes one step further and quantifies the \textit{rate} at which  $I(A:B_l)_{\rho^H}$ saturates when $l$ grows. 
Indeed if each size of $b_i$ and $l$ are sufficiently large, we have  
\begin{eqnarray*}
I(A:B_{l+1})_{\rho^H}- I(A: B_l)_{\rho^H} = I(A: b_{l+1} | B_{l})_{\rho^H} \leq C e^{- c\sqrt{l}},
\end{eqnarray*}
for some constants $C, c > 0$. 
Therefore, the mutual information of a 1D local Gibbs state saturates the upper bound of the area law at least subexponentially fast in $l$. 

\subsection{A short depth representation of 1D Gibbs states}

Theorem \ref{m.mthm2} ensures that there exist local CPTP-maps which approximately recover a 1D Gibbs state from tracing out operations on local regions. 
In other words, the Gibbs state can be prepared by local operations on reduced states on separated subregions. 
\bcor   \label{lowdepth preparation}
A Gibbs state of any 1D short-range Hamiltonian at constant (inverse) temperature $\beta>0$ can be well-approximated by a depth-two (mixed) circuit with each gate acting on $O(\log^2(n))$ qubits. 

In more detail, there is a CPTP-map (corresponding to the circuit) of the form 
\begin{equation*}
\Delta = \left( \bigotimes_i \Delta_{2, i} \right) \left( \bigotimes_i \Delta_{1, i} \right),
\end{equation*}
with each local CPTP-map $\Delta_{k, i}$ acting on $O(e^{O(\beta)} \log^2(n/\varepsilon))$ sites, with $\Delta_{k, i}$ and $\Delta_{k, j}$ acting on non-overlapping sites for $i \neq j$, such that
\begin{equation*}
\left \Vert \Delta(\tau) - \frac{e^{-\beta H}}{Z}  \right \Vert_1 \leq \varepsilon,
\end{equation*}
with $\tau$ the maximally mixed state. 
\ecor
The proof is given in Sec.~\ref{lowdeproof}. 
An earlier result \cite{hastings2006solving} proved that 1D Gibbs states of local Hamiltonians at finite constant temperature can be approximated by a matrix product operator of polynomial bond dimension, which implies
they can be constructed efficiently on a quantum computer. However this result does not give that the state can be constructed by a short depth quantum circuit, as 
Corollary \ref{lowdepth preparation} shows. A similar construction methods appear in Refs.~\cite{PhysRevB.94.155125, KastoryanoBrandao16} under certain assumptions on the approximate Markov property.

\begin{figure}[htbp]  
\begin{center}
\includegraphics[width=0.6\textwidth]{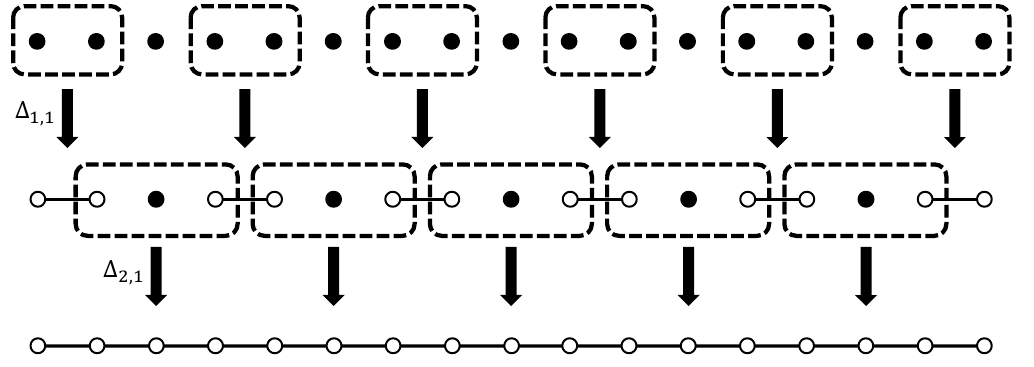}
\vspace{-5mm}
\end{center}
\caption{A schematic picture of the preparation algorithm for 1D Gibbs states. 
At the first step (upside), we perform CPTP-map $\bigotimes_i\Delta_{1,i}$ on a product state (black dots). 
Each $\Delta_{1,i}$ acts on a small set of spins (dotted circle) and produces the reduced state of the target Gibbs state on the set (connected dots). 
At the second step (downside), we perform another CPTP-map to concatenate these reduced states locally. 
Due to the approximate Markov property of the Gibbs state, the output state is close to the Gibbs state. }
\label{prep}
\end{figure}

\section{Gibbs State Representations of 1D States with the Approximate Markov Property}\label{sec:proof2}

In this section, we prove Theorem~\ref{mthm1} and Theorem~\ref{thm2} which states that quantum approximate Markov chains can be approximated by 1D local Gibbs states. We also prove Theorem~\ref{cmirelative}.
In these proofs, a  relationship between Gibbs states and the maximum entropy principle plays an important role. 
We first review the relationship and then turn to the proofs.  
\subsection{The maximum entropy state and Gibbs states}\label{sec:maxent}
The maximum entropy principle, introduced by E.\;T.\;Jaynes in classical statistical mechanics~\cite{PhysRev.106.620, PhysRev.108.171}, is a method to choose an inference under partial information represented by some linear constraints. 
According to the maximum entropy principle, the most \lq\lq{}unbiased\rq\rq{} inference is given by the probability distribution with maximum entropy among all distributions satisfying the linear constraints. The solution of this optimization problem is given by a Gibbs distribution, as can be shown by the method of Lagrange multipliers.  
 
This framework has been generalized to quantum systems (see, e.g., Refs.\;\cite{2010.5671W, weismaxent2015}). 
Especially we are interested in the case where the linear constraints are given by reduced density matrices. 
Let $\rho$ be a quantum state in $\cD(\cH_1\ot\cdots\ot\cH_n)$. Consider sets of subsystems labeled by $X_1,..,X_m$ with $X_i\subset\{1,...,n\}$ and let ${\bf X}=\{X_1,...,X_m\}$.  
We define the set $R_\rho({\bf X})$ by
\begin{equation}\label{rrdm}
R_\rho({\bf X}):=\left\{\sigma\in\cD(\cH_1\ot\cdots\ot\cH_n)|\sigma_{X_i}=\rho_{X_i}\,,\;(1\leq \forall i\leq m)\right\}\,.
\end{equation}
$R_\rho({\bf X})$ is the set of all states with the same reduced states as $\rho$ for all $X_i$. 
Since $R_\rho({\bf X})$ is a closed convex set, there exists a unique state such that
\begin{equation*}
\sigma_{\max}:=\argmax_{\sigma\in R_\rho(\bf X)}S(\sigma)\,.
\end{equation*}
We call $\sigma_{\max}$ {\it the maximum entropy state} in $R_\rho(\bf X)$. 
Similar to the classical setting, $\sigma_{\max}$ is given by a Gibbs state of a Hamiltonian with a specific structure.  
Let us consider the set of Gibbs states $E({\bf X})$ defined as
\begin{equation}\label{egibbs}
E({\bf X}):=\left\{\left.\frac{e^{-H}}{Z}\in\cD(\cH_1\ot\cdots\ot\cH_n)\,\right| H=\sum_{i=1}^{m}H_{X_i}\right\}\,,
\end{equation}
where $\text{supp}(H_{X_i})=X_i$. 
For any $\omega\in E({\bf X})$, it is proven  that the Pythagorean theorem 
\begin{equation}\label{Pytha}
S(\rho\|\omega)=S(\rho\|\sigma_{\max})+S(\sigma_{\max}\|\omega)\,
\end{equation}
holds (Corollary 3.7 \& Theorem 6.16,~\cite{2010.5671W}). The maximum entropy state $\sigma_{\max}$ is the unique element of the intersection of $R_\rho({\bf X})$ and $\overline {E({\bf X})}$: the closure of $E({\bf X})$ in the reverse-information topology~\cite{2010.5671W} defined as
\begin{equation*}
\overline {E({\bf X})}:=\left\{\sigma\in\cD(\cH_1\ot\cdots\ot\cH_n)\,\left| \inf_{\omega\in E({\bf X})}S(\sigma\|\omega)=0\right.\right\}\,.
\end{equation*} Therefore, the Pythagorean theorem~\eqref{Pytha} implies that
\begin{align}\label{maxent1}
\inf_{\omega\in {E({\bf X})}}S(\rho\|\omega)&=S(\rho\|\sigma_{\max})\,,
\end{align}
since $\inf_{\omega\in {E({\bf X})}}S(\sigma_{\max}\|\omega)=0$. 
By choosing $\omega$ in Eq.~\eqref{Pytha} as the completely mixed state, which is always contained in $E({\bf X})$, we have 
\begin{equation}\label{maxent2}
S(\rho\|\sigma_{\max})=S(\sigma_{\max})-S(\rho)\,
\end{equation}
and therefore
\begin{align}\label{relativedif}
\inf_{\omega\in {E({\bf X})}}S(\rho\|\omega)&=S(\sigma_{\max})-S(\rho)\,.
\end{align}
We will use these formulas in the proof of theorems. 

\subsection{Proof of Theorem~\ref{mthm1}}\label{subsecpthm1}

We now prove:\\
{\bf Theorem~\ref{mthm1}} 
{\it Let $\rho_{A_1 \ldots A_m}$ be a quantum $\varepsilon$-approximate Markov chain on a 1D open chain for $\veps>0$. Then there exists a short-range Hamiltonian $H = \sum_{i=1}^{m-1} h_{A_i, A_{i+1}}$ with $ \text{supp}(h_{A_i, A_{i+1}})=A_iA_{i+1}$, such that 
\begin{equation*}
S \left( \rho \left \Vert  \frac{e^{-H}}{Z} \right.  \right)\leq \varepsilon m. 
\end{equation*} 
\\
}
\begin{proof}

Let $\sigma_{A_1 \ldots A_m}$ be the maximum entropy state such that 
\begin{equation} \label{marginalsthesame}
\sigma_{A_{i} A_{i+1}} = \rho_{A_{i} A_{i+1}}
\end{equation}
for all $i \in [1, m-1]$.  We will show  
\begin{equation*}
S(\sigma) \leq S(\rho) + \varepsilon m.
\end{equation*}
The result then follows from Eq. (\ref{relativedif}), since  $\sigma_{A_1\ldots A_m}$ is an element of $\overline{E({\bf X})}$ for ${\bf X}=\{A_iA_{i+1}\}_{i=1}^{m-1}$ which is a (closure of) set of local Gibbs states. 

By strong subadditivity, we find
\begin{eqnarray}  \label{oneside}
S(A_1 \ldots A_m)_{\sigma} &\leq& S(A_1 A_2)_{\sigma} - S(A_2)_{\sigma} + S(A_2 \ldots A_m)_{\sigma} \nonumber \\
&\leq& S(A_1 A_2)_{\sigma} - S(A_2)_{\sigma} + S(A_2 A_3)_{\sigma} - S(A_3)_{\sigma} + S(A_3 \ldots A_m)_{\sigma} \nonumber \\
&\vdots& \nonumber \\
&\leq& \sum_{i=1}^{m-2} \left[S(A_i A_{i+1})_{\sigma} - S(A_{i+1})_{\sigma} \right] +S(A_{m-1}A_n)_\sigma\nonumber  \\
&=& \sum_{i=1}^{m-2} \left[ S(A_i A_{i+1})_{\rho} - S(A_{i+1})_{\rho} \right]+S(A_{m-1}A_m)_\rho
\end{eqnarray}
The last equality follows from Eq. (\ref{marginalsthesame}). Since $\rho_{A_1\ldots A_m}$ is a quantum $\veps$-approximate Markov chain, 
\begin{equation*}
I(A_i : A_{i+2}...A_m | A_{i+1})_\rho \leq \varepsilon \hspace{0.2cm} \forall i\in[1,m-2],
\end{equation*}
which can be rewritten as
\begin{equation*}
S(A_i A_{i+1})_\rho - S(A_{i+1})_\rho + S(A_{i+1}...A_m)_\rho \leq S(A_i A_{i+1}.... A_m)_\rho +\veps,
\end{equation*}
i.e., the strong subadditivity is saturated up to error $\varepsilon$. Therefore we obtain
\begin{eqnarray} \label{theotheriside}
\sum_{i=1}^{m-2} \left[S(A_i A_{i+1})_{\rho} - S(A_{i+1})_{\rho} \right] +S(A_{m-1}A_m)_\rho&\leq& \sum_{i=1}^{m-2} S(A_i |A_{i+1})_{\rho} + S(A_{m-2} A_{m-1} A_m) + \varepsilon \nonumber \\
 &\leq& \sum_{i=1}^{m-3} S(A_i |A_{i+1})_{\rho} + S(A_{m-3}\ldots A_m) + 2\varepsilon \nonumber \\
&\vdots& \nonumber \\
&\leq& S(A_1 |A_2)_{\rho} + S(A_2 \ldots A_m)_{\rho} + (m-2)\varepsilon \nonumber \\
&\leq& S(A_1...A_m)_{\rho} + (m-1)\varepsilon\,,  
\end{eqnarray}
where $S(A|B)_\rho:=S(AB)_\rho-S(B)_\rho$ is the conditional entropy. 
Combining Eqs. (\ref{oneside}) and (\ref{theotheriside}) we have
\begin{equation*}
S(\rho_A\|\sigma_A)=S(\sigma_A) - S(\rho_A) \leq (m-1)\veps\,.
\end{equation*}
Since $\sigma_A\in\overline{E({\bf X})}$, there exists a Gibbs state
\begin{equation*}
\omega_A = \frac{1}{Z} \exp \left(-\sum_{i=1}^{m-1} h_{A_i A_{i+1}} \right)\in E({\bf X})
\end{equation*}
which satisfies 
\begin{equation*}
S(\sigma_A\|\omega_A)\leq \varepsilon\,.
\end{equation*}
Using the Pythagorean theorem, we obtain 
\begin{equation*}
S(\rho_A\|\omega_A)=S(\rho_A\|\sigma_A)+S(\sigma_A\|\omega_A)\leq m\varepsilon\,,
\end{equation*}
which completes the proof. 
\end{proof}

\subsection{Proof of Theorem~\ref{thm2}} 
Let us first restate the theorem: \\
{\bf Theorem~\ref{thm2}} 
{\it Let $\rho_{A_1 \ldots A_m}$ be a quantum $\varepsilon$-approximate Markov chain on a 1D closed chain for $\veps>0$. 
\begin{itemize}
\item[$(i)$]  
Assume that $\rho_{A_1\ldots A_m}$ also satisfies 
\begin{equation*}
I(A_i:A\backslash A_{i-1}A_iA_{i+1})_\rho\leq\veps \hspace{0.2cm}\forall i\in[1,m]\,.
\end{equation*}
Then there exists a short-range Hamiltonian $H = \sum_{i} h_{A_{i-1},A_i, A_{i+1}}$, with $\text{supp}(h_{A_{i-1},A_i, A_{i+1}})=A_{i-1}A_i A_{i+1}$, such that 
\begin{equation*}
S \left( \rho \left \Vert  \frac{e^{-H}}{Z} \right.  \right)\leq \varepsilon m. 
\end{equation*} 

\item[$(ii)$]
Assume that for any $i\in[1,m]$, $\tr_{A_i}(\rho_{A_1\ldots A_m})$ is a quantum $\veps$-approximate Markov chain  for the 1D open chain $A_{i+1}A_{i+2}...A_{i-1}$ (we used $m+1\equiv 1$).  
Then there exists a short-range Hamiltonian $H = \sum_{i} h_{A_{i-1},A_i, A_{i+1}}$, with $\text{supp}(h_{A_{i-1},A_i, A_{i+1}})=A_{i-1}A_i A_{i+1}$, such that 
\begin{equation*}
S \left( \rho \left \Vert  \frac{e^{-H}}{Z} \right.  \right)\leq \varepsilon m. 
\end{equation*} 
\end{itemize}
}

\begin{proof}
Theorem~\ref{thm2}$(i)$ can be shown in the following way.  
Let $\sigma_{A_1 \ldots A_m}$ be the maximum entropy state such that 
\begin{equation*} 
\sigma_{A_{i-1}A_{i} A_{i+1}} = \rho_{A_{i-1}A_{i} A_{i+1}}
\end{equation*}
for all $i \in [1, m]$ (with the periodic boundary condition). 
As in the previous section, we recursively use the strong subadditivity and obtain 
\begin{align*}
S(A_1...A_{m})_\sigma&\leq S(A_1A_2A_3)_\sigma-S(A_1A_3)_\sigma+S(A_1A_3A_4...A_m)_\sigma\\
&\leq S(A_2|A_1A_3)_\sigma+S(A_3A_4A_5)_\sigma-S(A_3A_5)_\sigma+S(A_1A_3A_5...A_m)_\sigma\\
&\;\;\vdots\nonumber\\
&\leq \sum_{i=1}^{\lfloor\frac{m}{2}\rfloor}S(A_{2i}|A_{2i-1}A_{2i+1})_\sigma+S(A_1A_3A_5...A_{2{\lceil\frac{m}{2}\rceil}-1})_\sigma\\
&\leq\sum_{i=1}^{\lfloor\frac{m}{2}\rfloor}S(A_{2i}|A_{2i-1}A_{2i+1})_\sigma+\sum_{i=1}^{\lceil\frac{m}{2}\rceil}S(A_{2i-1})_\sigma\\
&=\sum_{i=1}^{\lfloor\frac{m}{2}\rfloor}S(A_{2i}|A_{2i-1}A_{2i+1})_\rho+\sum_{i=1}^{\lceil\frac{m}{2}\rceil}S(A_{2i-1})_\rho\,,
\end{align*}
where we used subadditivity $S(A_iA_j)\leq S(A_i)+S(A_j)$ in the second inequality. 
 From the additional assumption on the mutual information, 
\begin{equation*}
0\leq S(A_1)_\sigma+S(A_3)_\sigma-S(A_1A_3)_\sigma\leq I(A_1:A_3A_4...A_{m-1})_\sigma\leq\veps\,.
\end{equation*}
By using this type of inequalities, we obtain that
\begin{align*}
\sum_{i=1}^{\lceil\frac{m}{2}\rceil}S(A_{2i-1})_\rho&\leq S(A_1A_3)_\rho+\sum_{i=3}^{\lceil\frac{m}{2}\rceil}S(A_{2i-1})_\rho+\veps\\
&\;\;\vdots\nonumber\\
&\leq S(A_1A_3A_5...A_{2\lceil\frac{m}{2}\rceil-1})_\rho+\left(\left\lceil\frac{m}{2}\right\rceil-1\right)\veps\,.
\end{align*}
Since $\rho_{A_1...A_n}$ is a quantum $\veps$-approximate Markov chain on a closed chain, one can further show that
\begin{align*}
S(A_1...A_{m})_\sigma&\leq\sum_{i=1}^{\lfloor\frac{m}{2}\rfloor}S(A_{2i}|A_{2i-1}A_{2i+1})_\rho+S(A_1A_3A_5...A_{2\lceil\frac{m}{2}\rceil-1})_\rho+\left(\left\lceil\frac{m}{2}\right\rceil-1\right)\veps\\
&\leq S(A_1A_2...A_{m})_\rho+(m-1)\veps\,.
\end{align*}
The rest part of the proof is the same as the proof of Theorem~\ref{mthm1}. 

To prove Theorem~\ref{thm2}$(ii)$, we use the strong subadditivity to the maximum entropy state $\sigma_{A_1\ldots A_m}$ to obtain: 
\begin{align*}
S(A_1...A_{m})_\sigma&\leq S(A_mA_1A_2)_\sigma-S(A_mA_2)_\sigma+S(A_2A_3...A_m)_\sigma\\
&\leq S(A_1|A_mA_2)_\sigma+S(A_2A_3)_\sigma-S(A_3)_\sigma+S(A_3...A_m)_\sigma\\
&\;\;\vdots\nonumber\\
&\leq S(A_1|A_mA_2)_\sigma+\sum_{i=2}^{m-2}S(A_i|A_{i+1})_\sigma+S(A_{m-1}A_m)_\sigma\\
&\leq S(A_1|A_mA_2)_\rho+\sum_{i=2}^{m-2}S(A_i|A_{i+1})_\rho+S(A_{m-1}A_m)_\rho\,.
\end{align*}
Recall that $\rho_{A_2...A_m}$ is a quantum $\veps$-approximate Markov chain by assumption. Therefore, we have 
\begin{align*}
S(A_1|A_mA_2)_\rho+\sum_{i=2}^{m-2}S(A_i|A_{i+1})_\rho+S(A_{m-1}A_m)_\rho&\leq S(A_1|A_mA_2)_\rho+S(A_2...A_m)_\rho+(m-2)\veps\nonumber\\
&\leq S(A_1...A_m)_\rho+(m-1)\veps\,,
\end{align*}
from which we complete the proof in the same way as in the proof of Theorem~\ref{mthm1}. 
\end{proof}
\subsection{Proof of Theorem~\ref{cmirelative}}
Finally, we prove Theorem~\ref{cmirelative}: 

{\bf Theorem}~\ref{cmirelative}. {\it
Consider a 1D spin chain $X=X_1X_2...X_m$ with the size $N=|X_1...X_m|$. Let $\rho_{X_1...X_m}$ be a state such that  the reduced state obtained by tracing out $X_i$ is a quantum $\varepsilon$-approximate Markov chain  for all $i\in[1, m]$.   
Define the set of Gibbs states of short-range Hamiltonians with interaction strength $K$ as 
\begin{equation*}
E^K_{nn}:=\left\{\frac{1}{Z}e^{-H}\left|\; H=\sum_ih_{X_iX_{i+1}}, \;\|h_{X_iX_{i+1}}\|\leq K\right.\right\}\,.
\end{equation*}
Then, for $K=\Theta(N)$ and sufficiently small $\veps>0$, there exists a constant $c>0$ such that  for any tripartition $ABC$ of the whole system such that $B$ shields $A$ from $C$, it holds that
\begin{equation*}
\min_{\mu\in E^K}S\left(\rho_{X}\left\| \mu\right.\right)=I(A:C|B)_\rho +\epsilon(N,\delta)
\end{equation*}
and
\begin{equation*}
\left|\epsilon(N,\delta)\right|\leq cN^{\frac{5}{2}}\delta^{\frac{1}{16}}\,,
\end{equation*}   
where $\delta=8\sqrt{\veps}+2^{-N}$.
}

The strategy of the proof is as follows. We first construct a global state $\trho'_{ABC}$ on $ABC$ from a reduced state of $\rho_{ABC}$ by using recovery maps.  We then introduce $\trho_{ABC}$, a modification of $\trho'_{ABC}$ and define a Gibbs state of a nearest-neighbor Hamiltonian $\tpi_{ABC}$ from its reduced states. Finally we show that this Gibbs state is almost the closest state in the set of Gibbs states of bounded nearest-neighbor Hamilntonians. 
\begin{proof}
Without loss of generality, we consider a system defined on a 1D closed spin chain, e.g., $X_1\equiv X_{m+1}$, and assume that $|X_1|=\max_i|X_i|$. 
Let $A\equiv X_1$, $B_1\equiv X_mX_2$, $B_2\equiv X_{m-1}X_3$ and $C\equiv X_4X_5\ldots X_{m-2}$. 
In the following, we use both notations $X_1\ldots X_m$ and $ABC$ interchangeably. 
Note that if we choose another tripartition $A\rq{}B\rq{}C\rq{}$ satisfying the condition instead of $ABC$, the chain rule of the conditional mutual information:
\begin{equation}\label{eq:chainrule}
I(A:CD|B)_\rho=I(A:C|B)_\rho+I(A:D|BC)_\rho
\end{equation}
and the assumption imply that	
\begin{equation*}
\left|I(A:C|B)_\rho-I(A\rq{}:C\rq{}|B\rq{})_\rho\right|\leq\veps\,.
\end{equation*}

From the results in Ref.~\cite{Fawzi2015}, there exist CPTP maps called (approximate) recovery maps such that
\begin{align}
\left\|\rho_{AB_1B_2}-\Lambda_{B_1\to AB_1}(\rho_B)\right\|_1&\leq2\sqrt{\epsilon}\,,\label{eq1p1}\\
\left\|\rho_{B_1B_2C}-\Lambda_{B_2\to B_2C}(\rho_B)\right\|_1&\leq2\sqrt{\epsilon}\label{eq2p1}\,,
\end{align} 
where we omitted the identity maps for simplicity. 
By using these maps, we define a global state
\begin{equation*}
{\trho}\rq{}_{ABC}:=\Lambda_{B_2\to B_2C}\circ\Lambda_{B_1\to AB_1}(\rho_{B})\,.
\end{equation*}
Since the CPTP-maps recover reduced states with good accuracy, $\rho_{ABC}\rq{}$ has almost same bipartite marginals on, e.g., $AB$. 
\begin{align}
\hspace{-3mm}\|\rho_{AB}-\trho\rq{}_{AB}\|_1&\leq\left\|\rho_{AB}-\Lambda_{B_1\to AB_1}(\rho_B)\right\|_1
+\left\|\Lambda_{B_1\to AB_1}(\rho_{B})-\trho_{AB}\rq{}\right\|_1\nonumber\\
&\leq\left\|\rho_{AB}-\Lambda_{B_1\to AB_1}(\rho_B)\right\|_1+\left\|\Lambda_{B_1\to AB_1}(\rho_{BC})-\trho_{ABC}\rq{}\right\|_1\nonumber\\
&\leq2\sqrt{\veps}+\left\|\Lambda_{B_1\to AB_1}(\rho_{BC})-\Lambda_{B_1\to AB_1}\circ\Lambda_{B_2\to B_2C}(\rho_{B})\right\|_1\nonumber\\
&\leq4\sqrt{\veps}\label{eq2.5p1}.
\end{align}
The first inequality follows from the triangle inequality of the trace norm, and the second inequality follows from the monotonicity of the trace-norm under $\tr_C$. 
The same bound holds for marginals on $BC$ as well. 
Eq.~\eqref{eq1p1} and Eq.\;\eqref{eq2.5p1} imply that $\trho_{ABC}\rq{}$ is approximately recoverable state:  
\begin{equation}\label{eq3p1}
\left\|\trho_{ABC}\rq{}-\Lambda_{B_2\to B_2C}(\trho_{AB}\rq{})\right\|_1\leq \left\|\trho_{ABC}\rq{}-\Lambda_{B_2\to B_2C}(\rho_{AB})\right\|_1+\left\|\rho_{AB}-\trho_{AB}\rq{}\right\|_1\leq6\sqrt{\veps}\,.
\end{equation}
For two quantum states $\rho_{AB}, \sigma_{AB}$ satisfying $\|\rho_{AB}-\sigma_{AB}\|_1<\delta\leq1$, the (Alicki-)Fannes inequality~\cite{AF04} 
\begin{align*}
\left|S(A|B)_\rho-S(A|B)_\sigma\right|&\leq4\delta|A|+2h_2(\delta) \\
&\leq6\sqrt{\delta}|A|\,
\end{align*}
holds with the binary entropy $h_2(\delta)=H(p=\{\delta,1-\delta\})$. 
This inequality and Eq.~\eqref{eq3p1} yields 
\begin{align*}
I(A:C|B)_{\trho\rq{}}&=S(A|B)_{\trho'}-S(A|BC)_{\trho'}\\
&\leq S(A|BC)_{\trho''}-S(A|BC)_{\trho'}\\
&\leq 24\sqrt{\veps}|A|+2h_2\left(6\sqrt{\veps}\right)\;\\
&\leq6\sqrt{6}|A|\veps^{\frac{1}{4}}\,,
\end{align*}
where we denote $\trho''_{ABC}:=\Lambda_{B_2\to B_2C}(\trho_{AB}\rq{})$ in the second line which follows from the data processing inequality for $\Lambda_{B_2 \to B_2C}$. 
Therefore, $\trho_{ABC}'$ is a quantum approximate Markov chain for small $\veps>0$. 

Next, define a full-rank modification of $\trho_{ABC}\rq{}$, that is,
\begin{equation}\label{defsigma}
\trho_{ABC}:=\left(1-\frac{1}{2^{N-1}}\right)\trho_{ABC}\rq{}+\frac{1}{2^{N-1}}\tau_{ABC}\,,
\end{equation}
where $\tau_{ABC}$ is the completely mixed state on $ABC$ (recall that $N$ represents the logarithm of the total dimension). Since by definition 
\begin{equation}\label{eq3.5p1}
\|\trho_{ABC}-\trho_{ABC}\rq{}\|_1\leq2^{-N}\,,
\end{equation} 
the Fannes inequality implies that 
\begin{align}
\hspace{-3mm}I(A:C|B)_\trho&\leq I(A:C|B)_{\trho\rq{}}+ \frac{12|A|}{\sqrt{2^N}}\label{eq3.9p1}\\
\label{eq4p1}&\leq 6\sqrt{6}|A|\veps^{\frac{1}{4}}+ \frac{12|A|}{\sqrt{2^N}}\,.
\end{align}
Therefore $\trho_{ABC}$ is still an approximate Markov chain for large $N$. 

Then, we construct a Gibbs state $\tpi_X$ 
\begin{equation}\label{tpidef}
\tpi_X:=\frac{1}{Z}e^{-H^\trho_X}\,
\end{equation}
from the reduced states of $\trho_{ABC}$, where $Z$ is the normalizer and
\begin{equation*}
H^\trho_X:=-\sum_{i=1}^m\left(\ln\trho_{X_iX_{i+1}}-\ln\trho_{X_i}\right)\,.
\end{equation*}
$\tpi_X$ is an element of $E^K_{nn}$ with $K=\Th(N)$. In the following, we show that $\tpi$ is close to $\trho$.

By definition of $H_X^\trho$, it holds that
\begin{align}
S(\trho_X\|e^{-H^\trho_X})&=\sum_{i=1}^mS(X_{i}|X_{i+1})_\trho-S(X_1\ldots X_m)_\trho\,.\label{nagaiyo}
\end{align}
Note that $\sum_{i=1}^mS(X_{i}|X_{i+1})_{\trho}=\sum_{i=1}^mS(X_{i+1}|X_{i})_{\trho}$ by $X_{m+1}=X_1$.
By an iterative calculation, we have
\begin{align*}
\sum_{i=1}^mS(X_{i}|X_{i+1})_{\trho}
&=S(X_1|X_2)_{\trho}+S(X_2|X_3)_{\trho}+S(X_3|X_4)_{\trho}+\sum_{i=4}^{m}S(X_i|X_{i+1})_{\trho}\\
&=S(X_1|X_2)_{\trho}+I(X_2:X_4|X_3)_{\trho}+S(X_2X_3|X_4)_{\trho}+\sum_{i=4}^{m}S(X_i|X_{i+1})_{\trho}\\
&\;\;\vdots\nonumber\\
&=S(X_1|X_2)_{\trho}+S(X_2...X_{m-1}|X_m)_{\trho}+S(X_m|X_1)_{\trho}\nonumber\\&\quad+\sum_{i=3}^{m-1}I(X_2...X_{i-1}:X_{i+1}|X_{i})_{\trho}\\
&=S(X_2\ldots X_{m-1}|X_m)_{\trho}-S(X_2)_{\trho}+S(X_mX_1X_2)_{\trho}\nonumber\\&\quad+I(X_m:X_2|X_1)_{\trho}+\sum_{i=3}^{m-1}I(X_2...X_{i-1}:X_{i+1}|X_{i})_{\trho}\,.
\end{align*}
By using the subadditivity $S(X_2X_m)\leq S(X_2)+S(X_m)$, we obtain that
\begin{align*}
\eqref{nagaiyo}\leq I(X_1:X_3... X_{m-1}|X_2X_m)_{\trho}+I(X_m:X_2|X_1)_{\trho}+\sum_{i=3}^{m-1}I(X_2...X_{i-1}:X_{i+1}|X_{i})_{\trho}\,.
\end{align*}
Therefore,  we have
\begin{align}
S\left(\trho_X\left\|e^{-H^\trho_X}\right.\right)&\leq I(X_m:X_2|X_1)_{\trho}
+\sum_{i=2}^{m-2}I(X_2...X_i:X_{i+2}|X_{i+1})_{\trho}\nonumber\\
&\qquad+I(A:B_2|B_1)_{\trho}+I(A:C|B_1B_2)_{\trho}\,,\label{cmitakusan}
\end{align}
where we used the chain rule~\eqref{eq:chainrule}. The first three terms only depend on marginals of $\trho_{ABC}$ on $AB$ or $BC$. 
Since $\trho_{ABC}$ is close to $\trho'_{ABC}$~\eqref{eq3.5p1}, $\trho_{ABC}$ also has marginals close to $\rho_{ABC}$ on these regions:  
\begin{equation*}
\left\|\trho_{AB}-\rho_{AB}\right\|_1, \left\|\trho_{BC}-\rho_{BC}\right\|_1\leq 8\sqrt{\veps}+2^{-N}\equiv\delta\,.
\end{equation*}
Thus, as in Eq.~\eqref{eq3.9p1}, the Fannes inequality implies that
\begin{align*}
 I(X_m:X_2|X_1)_{\trho}+\sum_{i=2}^{m-2}I(X_2...X_i:X_{i+2}|X_{i+1})_{\trho}&\leq I(X_m:X_2|X_1)_{\rho}+\sum_{i=2}^{m-2}I(X_2...X_i:X_{i+2}|X_{i+1})_{\rho}\nonumber\\
 &\quad+12\left(|X_2|+\sum_{i=2}^{m-2}|X_{i+2}|\right)\sqrt{\delta}\\
&\leq (m-1)\veps+12(m-2)|X_1|\sqrt{\delta}\,.
\end{align*}
Combining with Eq.~\eqref{eq4p1}, we obtain
\begin{align*}
S\left(\trho_X\left\|e^{-H^\trho_X}\right.\right)&\leq (m-1)\veps+12(m-2)|X_1|\sqrt{\delta}+ 6\sqrt{6}|X_1|\veps^{\frac{1}{4}}+ \frac{12|X_1|}{\sqrt{2^N}}\\
&\leq\cO\left(N\sqrt{\delta}\right)\,,
\end{align*}
where we used $m|X_1|=\Th(N)$ and the asymptotic notation $\cO(f(N,\veps))$ as $N\to\infty$ and $\veps\to0$ (and therefore $\delta\to0$).

Let us estimate the partition function $Z=\tr e^{-H^{\trho}_X}$. 
By Pinsker inequality, the above bound implies
\begin{align}
\left|\tr\left(e^{-H^\trho_X}\right)-1\right|&\leq \left\|\trho_X-e^{-H^\trho_X}\right\|_1\nonumber\\
&\leq\sqrt{2 S\left(\trho_X\left\|e^{-H^\trho_X}\right.\right)}\nonumber\\
&\leq\cO\left(N^{\frac{1}{2}}\delta^{\frac{1}{4}}\right)\,.\label{eq:uuperbb}
\end{align}
where we used the inequality $\left|\|A\|_1-\|B\|_1\right|\leq\|A-B\|_1$.  
For given $N$, we assume $\delta$ is sufficiently small so that Eq.~\eqref{eq:uuperbb} is smaller than 1. We then obtain
\begin{equation*}
|\log Z|\leq\cO\left(N^{\frac{1}{2}}\delta^{\frac{1}{4}}\right)\,. 
\end{equation*}
Thus, the difference between $\trho_X$ and $\tpi_X$ is bounded as
\begin{align}
\left\|\trho_X-\tpi_X\right\|_1&\leq\sqrt{2S(\trho_X\|\tpi_X)}\nonumber\\
&=\sqrt{2\left(S(\trho_X\|e^{-H^\trho_X})-\log Z\right)}\nonumber\\
&\leq\cO\left(N^{\frac{1}{2}}\delta^{\frac{1}{8}}\right)\,.\label{eq6p1}
\end{align}
Again, by the Fannes inequality, the conditional mutual information of $\tpi_X$ is bounded as
\begin{equation*}
I(A:C|B)_\tpi\leq\cO\left(|X_1|N^{\frac{1}{4}}\delta^{\frac{1}{16}}\right)\,.
\end{equation*}
The marginal of $\tpi$ on $AB$ satisfy
\begin{align}
\|\rho_{AB}-\tpi_{AB}\|_1&\leq\|\rho_{AB}-\trho\rq{}_{AB}\|_1+\|\trho\rq{}_{AB}-\trho_{AB}\|_1+\|\trho_{AB}-\tpi_{AB}\|_1\nonumber\\
&\leq\cO\left(N^{\frac{1}{2}}\delta^{\frac{1}{8}}\right).\label{eq6.55p1}
\end{align}
Here, we used Eq.~\eqref{eq2.5p1}, Eq.~\eqref{eq3.5p1} and Eq.~\eqref{eq6p1} in the second inequality. In the same way, we also obtain
\begin{equation}\label{eq6.5p1}
\|\rho_{BC}-\tpi_{BC}\|_1\leq\cO(N^{\frac{1}{2}}\delta^{\frac{1}{8}})\,.
\end{equation}

Finally, we show that $I(A:C|B)_\rho$ approximates the distance between $\rho$ and the set of Gibbs states $E_{nn}^K$ in terms of the relative entropy. 
By combining Eq.~\eqref{eq6.55p1} and Eq.~\eqref{eq6.5p1} with the Fannes inequality, we have that
\begin{align}
&\left|I(A:C|B)_\rho-\left(S(ABC)_\tpi-S(ABC)_\rho\right)\right|\nonumber\\
&\leq \left|S(A|B)_\rho+S(BC)_\rho-S(ABC)_\tpi\right|\nonumber\\
&\leq\left|S(A|B)_\rho-S(A|B)_\tpi\right|+\left|S(BC)_\rho-S(BC)_\tpi\right|+I(A:C|B)_\tpi\nonumber\\
&\leq\cO\left(N^{\frac{5}{4}}\delta^{\frac{1}{16}}\right)\,.\label{eq99p1}
\end{align}
By definition, it holds that
\begin{equation*}
\min_{\mu\in E_{nn}^K}S(\rho_{ABC}\|\mu_{ABC})=S(ABC)_\rho-\min_{\mu\in E_{nn}^K}\tr(\rho\log\mu)\,.
\end{equation*}
Here, $\log\mu\propto H_{AB}+H_{BC}$ for some bounded Hermitian operators $H_{AB}$ and $H_{BC}$ satisfying $\|H_{AB}\|+\|H_{BC}\|\leq\cO(mK)$. 
Eq.~\eqref{eq6.55p1} and Eq.~\eqref{eq6.5p1} yield that
\begin{align}
\left|\tr(\rho_Y O_Y)-\tr(\tpi_YO_Y)\right|&\leq\|O\|\|\rho_Y-\tpi_Y\|\nonumber\\
&\leq\cO\left(\|O\|N^{\frac{1}{2}}\delta^{\frac{1}{8}}\right)\label{yoiy}
\end{align}
for $Y=AB,BC$. Therefore, we can approximate $\tr(\rho\log\mu)$ by $\tr(\tpi\log\mu)$, and we have
\begin{align*}
\min_{\mu\in E_{nn}^K}S(\rho_{ABC}\|\mu_{ABC})&=S(ABC)_\rho-\min_{\mu\in E_{nn}^K}\tr(\tpi\log\mu)+\cO\left(mKN^{\frac{1}{2}}\delta^{\frac{1}{8}}\right)\\
&=S(ABC)_\tpi-\min_{\mu\in E_{nn}^K}\tr(\tpi\log\mu)+I(A:C|B)_\rho\nonumber\\
&\quad+\cO\left(N^{\frac{5}{4}}\delta^{\frac{1}{16}}\right)+\cO\left(N^{\frac{5}{2}}\delta^{\frac{1}{8}}\right)\\
&=\min_{\mu\in \cE_{nn}^K}S\left(\tpi_{ABC}\left\|\mu_{ABC}\right.\right)+I(A:C|B)_\rho+\cO\left(N^{\frac{5}{2}}\delta^{\frac{1}{16}}\right)\\
&=I(A:C|B)_\rho+\cO\left(N^{\frac{5}{2}}\delta^{\frac{1}{16}}\right)\,,
\end{align*}
where we used $mK=\Th(N^2)$ and Eq.~\eqref{eq99p1} in the secondline. The third inequality follows from Eq.~\eqref{eq99p1} and the last line follows from $\tpi\in E_{nn}^K$. 
\end{proof}
\section{1D Quantum Gibbs states are Quantum Approximate Markov Chain}\label{sec:proof}
In this section, we provide a proof of Theorem~\ref{m.mthm2}, Corollary~\ref{cor1} and Corollary~\ref{lowdepth preparation}. 
A key point of the proof is that if a short-range Hamiltonian changes locally, the corresponding Gibbs state also changes quasi-locally. 
To obtain operators representing changes of the Gibbs state, we employ quantum belief propagation equations which have been studied in Refs.~\cite{PhysRevB.76.201102, PhysRevB.86.245116}. 
We first introduce these technical tools, and then show the proofs.  
We discuss another approach to prove Theorem~\ref{m.mthm2} based on the results of Ref.~\cite{Araki1969} in Appendix $A$. 
\subsection{Perturbative analysis of Gibbs states}
Let us consider an one-parameter family of a Hamiltonian  $H$ on a spin lattice with a perturbation operator $V$
\begin{equation}
H(s)=H+sV\,,
\end{equation}
where $s\in[0,1]$. 
The change of the corresponding Gibbs state due to a small change of $s$ can be computed through a quantum belief propagation equation~\cite{PhysRevB.76.201102, PhysRevB.86.245116}:
\begin{equation}\label{eq:derivativeH}
\frac{d}{ds}e^{-\beta H(s)}=-\frac{\beta}{2}\left\{e^{-\beta H(s)},\Phi_\beta^{H(s)}(V)\right\}\,,
\end{equation}
where the operator  $\Phi_\beta^{H(s)}(V)$ is given by~\cite{PhysRevB.86.245116}
\begin{equation}
\Phi_\beta^{H(s)}(V)_{ij}:=V_{ij}{\tilde f}_\beta(E_i(s)-E_j(s))\,
\end{equation}
in the energy eigenbasis of $H(s)=\sum_iE_i(s)|i\rangle\langle i|$~\footnote{Each $|i\>$ depends on $s$ as well as the eigenvalues $E_i(s)$}, with ${\tilde f}_\beta(\omega)=\frac{{\rm tanh}(\beta\omega/2)}{\beta\omega/2}$. 
Using the Fourier transform $f_\beta(t)=\frac{1}{2\pi}\int d\omega{\tilde f}_\beta(\omega)e^{i\omega t}$, 
$\Phi_\beta^{H(s)}(V)$ can be written as the integral form:  
\begin{equation}
\Phi_\beta^{H(s)}(V)=\int_{-\infty}^\infty dt f_\beta(t)e^{-iH(s)t}Ve^{iH(s)t}\,.
\end{equation}
Taking the formal integration of Eq.~\eqref{eq:derivativeH}, we obtain
\begin{equation}\label{eq:int}
e^{-\beta H(1)}=Oe^{-\beta H(0)}O^\dagger\,,
\end{equation}
where the operator $O$ is defined as
\begin{align*}
O&={\cal T}\exp\left[-\frac{\beta}{2}{\int_0^1}ds\rq{}\Phi_\beta^{H(s\rq{})}(V)\right]\\
&=\sum_{n=0}^\infty\left(-\frac{\beta}{2}\right)^n\int_0^1ds\rq{}_1\int_0^{s\rq{}_1}ds\rq{}_2\ldots\int_0^{s\rq{}_{n-1}}\hspace{-0.25cm}ds\rq{}_n \Phi_\beta^{H(s\rq{}_n)}(V)\ldots \Phi_\beta^{H(s\rq{}_1)}(V)\,,
\end{align*}
with ${\cal T}$ the time-ordering operation. 
Since we have $dtf_\beta(t)=\frac{dt}{\beta}f_1(\frac{t}{\beta})$, it holds that 
\begin{align}
\left\|\Phi_\beta^{H(s)}(V)\right\|&=\left\|\int_{-\infty}^\infty dt\rq{} f_1(t\rq{})e^{-i\beta H(s)t\rq{}}Ve^{i\beta H(s)t\rq{}}\right\|\nonumber\\
&\leq\|V\|\left|\int_{-\infty}^\infty dt\rq{} f_1(t\rq{})\right|\nonumber\\
&=\|V\|\,.\label{upbphiV}
\end{align}
The integral in the last equality can be calculated through the series expansion:
\begin{equation}
\frac{{\rm tanh}(x)}{x}=\sum_{k=0}^\infty \frac{2}{x^2+\left(k+\frac{1}{2}\right)^2\pi^2}\,.
\end{equation}
The upper bound of $\|\Phi_\beta^{H(s)}(V)\|$ implies the upper bound of $\|O\|$ that is given by
\begin{equation}\label{boundO}
\|O\|\leq e^{\frac{\beta}{2}\|V\|}\,.
\end{equation}

When the Hamiltonian defined on a many-body system is short-ranged, time-evolutions of a local operator is restricted by using the Lieb-Robinson bound~\cite{lieb1972}.
Suppose that $H$ is a Hamiltonian obeying the Lieb-Robinson bound, and $O_A$ and $O_B$ are observables supported on local regions $A$ and $B$, respectively. 
Then, the Lieb-Robinson bound for these operators is formulated as 
\begin{equation}
\left\|\left[O_A, e^{-iHt}O_Be^{iHt}\right]\right\|\leq c\|O_A\|\|O_B\|\min(|A|,|B|)e^{c\rq{}(vt-d(A,B))}\,,
\end{equation}
where $c,v\geq0$, $c\rq{}>0$ are constants. 
Assume that $H(0)$ is a short-range Hamiltonian and $supp(V)$ is a simply-connected local region. Then $H(s)$ obeys the Lieb-Robinson bound for all $s\in[0,1]$. 
Since $f_\beta(t)$ decays fast in $|t|$, the Lieb-Robinson bound implies that the effective support of $\Phi_\beta^{H(s)}(V)$ can be restricted to a local region $\cV_l$, which contains all sites within distance $l$ from $supp(V)$. 
More precisely, there exist positive constants $c\rq{} $ and $v$, which is determined by $H(s)$, such that~\cite{PhysRevB.86.245116}
\begin{equation}\label{eq:restriction1}
\left\|\Phi_\beta^{H(s)}(V)-\tr_{\cV_l^c}\left[\Phi_\beta^{H(s)}(V)\right]\ot\frac{1}{d_{\cV_l^c}}\idty_{\cV_l^c}\right\|\leq c\rq{}\|V\|e^{-\frac{c\rq{}l}{1+c\rq{}v\beta/\pi}}\,,
\end{equation}
where $d_{\cV_l^c}:=\dim\cH_{\cV_l^c}$. 
We also define the integral of the restricted operator in Eq.~\eqref{eq:restriction1} as 
\begin{equation}\label{eq:restriction2}
O_{\cV_l}:={\cal T}\exp\left(-\frac{\beta}{2}{\int_0^1}ds\rq{}\tr_{\cV_l^c}\left[\Phi_\beta^{H(s\rq{})}(V)\right]\ot\frac{1}{d_{\cV_l^c}}\idty_{\cV_l^c}\right)\,.
\end{equation}
This operator is also localized on $\cV_l$ and approximates $O$ with good accuracy. 
Let us choose $c\rq{}$ and $v$ so that Eq.~\eqref{eq:restriction1} holds for all $s\in[0,1]$. Then, we obtain that
\begin{equation}\label{eq:defOs}
\|O-O_{\cV_l}\|\leq \frac{c\rq{}\beta\|V\|}{2}e^{\frac{(1+c\rq{})\beta\|V\|}{2}}e^{-\frac{c\rq{}l}{1+c\rq{}v\beta/\pi}}.
\end{equation}
To see this, consider some parametrized operators $Q(s)$ and ${\tilde Q}(s)$ satisfying $\|Q(s)\|,\|{\tilde Q}(s)\|\leq C$ and $\|Q(s)-{\tilde Q}(s)\|\leq \Delta$ for all $s\in[0,1]$. From the simple calculation, we obtain
\begin{align*} 
Q(s_n)Q(s_{n-1})\cdots Q(s_1)&={\tilde Q}(s_n){\tilde Q}(s_{n-1})\cdots {\tilde Q}(s_1)\nonumber\\
&\quad+\sum_{j=1}^nQ(s_n)\cdots Q(s_{j+1})\Delta_j{\tilde Q}(s_{j-1})\cdots {\tilde Q}(s_1)\,
\end{align*}
where $\Delta_j=Q(s_j)-{\tilde Q}(s_j)$ and $\|\Delta_j\|\leq\Delta$. Therefore, we obtain that
\begin{equation}\label{eq:boundqq}
\|Q(s_n)Q(s_{n-1})\cdots Q(s_1)-{\tilde Q}(s_n){\tilde Q}(s_{n-1})\cdots {\tilde Q}(s_1)\|\leq nC^{n-1}\Delta\,.
\end{equation}
In our case, $Q(s)=\Phi_\beta^{H(s)}(V)$ and ${\tilde Q}(s)=\tr_{\cV_l^c}\left[\Phi_\beta^{H(s)}(V)\right]\ot\frac{1}{d_{\cV_l^c}}\idty_{\cV_l^c}$. 
We can choose  $\Delta=c\rq{}e^{-\frac{c\rq{}l}{1+c\rq{}v\beta/\pi}}\|V\|$. 
From Eq.~\eqref{upbphiV} and Eq.~\eqref{eq:restriction1}, their norms can be bounded as
\begin{align*}
\|Q(s)\|, \|{\tilde Q}(s)\|&\leq\|Q(s)\|+\|{\tilde Q}(s)-Q(s)\|\\
&\leq \left(1+c\rq{}e^{-\frac{c\rq{}l}{1+c\rq{}v\beta/\pi}}\right)\|V\|\\
&\leq (1+c\rq{})\|V\|\,.
\end{align*}
Therefore, Eq.~\eqref{eq:boundqq} holds for $C=(1+c\rq{})\|V\|$. By inserting Eq.~\eqref{eq:boundqq} to the definition of $O$, we obtain Eq.~\eqref{eq:defOs}.

\subsection{Proof of Theorem~\ref{m.mthm2}}\label{sec:prth1}
For the convenience of the reader, we restate Theorem~\ref{m.mthm2} below. \\\\
{\bf Theorem~\ref{m.mthm2}}
{\it 
Let $H = \sum_i h_i$ be a short-range 1D Hamiltonian with $\Vert h_i \Vert \leq 1$ and $l_0, C, c>0$ be universal constants. For an inverse temperature $\beta > 0$ and any partition $ABC$ with $d(A, C)\geq l_0$,  there exists a CPTP-map $\Lambda_{B\to BC} : \cD({\cal H}_B) \rightarrow  \cD({\cal H}_B  \otimes  {\cal H}_C)$ such that
\begin{equation}
\left \Vert  \rho^{H_{ABC}}  -\Lambda_{B\to BC}(\rho^{H_{ABC}}_{AB}) \right \Vert_1 \leq e^{- q(\beta) \sqrt{d(A,C)}}\,,\nonumber
\end{equation}
where $q(\beta) =  ce^{- c' \beta }$ if the correlation length of $\rho^{H_{ABC}}$ is $\xi=e^{(\cO(\beta)}$. 
}

The proof of Theorem~\ref{m.mthm2} consists of three steps. In the first step, we show that there exists a CP-map which recovers the Gibbs state from 
the reduced state with exponentially good accuracy  (Lemma~\ref{lem1}). In the second, we normalize the CP-map to make it trace non-increasing  
i.e. we show the existence of a quantum operation which succeed to recover the Gibbs state with some probability (Lemma~\ref{l.lemma2}). 
We also show that the success probability is a constant of system size. Finally, we construct a CPTP-map from the probabilistic operation by employing 
a repeat-until-success strategy in the third step. 

Let us begin with the following lemma. 
\blm\label{lem1}
For any 1D Gibbs state $\rho^{H_{ABC}}$ of a short-range Hamiltonian on a system with a partition $ABC$ such that $l:=d(A,C)/2>r$, there exists a CP map $\kappa_{B\to BC}$, non-negative constants $c\rq{}$ and $v$ such that
\begin{equation}\label{lemma1}
\|\rho^{H_{ABC}}-\kappa_{B\to BC}(\rho^{H_{ABC}}_{AB})\|_1 \leq C_1(\beta)e^{-q_1(\beta)l}\,,
\end{equation}
where $C_1(\beta)$ is a non-negative constant and $q_1(\beta)=\frac{c\rq{}}{1+c\rq{}v\beta/\pi}$, as defined in Eq.~\eqref{eq:restriction1}.
\elm
\begin{proof} 
Let us consider a short-range Hamiltonian $H=\sum_{i}h_{i}$ with the range $r$. 
Without loss of generality, we introduce a tripartition $ABC$ of a 1D system so that each subsystem is simply connected and $d(A,C)$ is chosen to be $2l$ for some integer $l>r$. We then split region $B$ into the left half $B^L$, which touches $A$, and the right half $B^R$ which touches $C$ (Fig.~\ref{bh}). We denote the sum of the all interactions $h_{i}$ acting on both $B^L$ and $B^R$ by $H_{B^M}=\sum_{j:i\in{{\rm supp}}(h_j)}h_{j}$. By assumption, $\|H_{B^M}\|\leq J$ for some constant $J\geq0$.   
\brem When $B$ consists of a fixed number of simply connected regions, each connected component neighboring both $A$ and $C$ is divided into two halves as in the same way. Then, $H_{B^M}$ is the sum of all interaction terms acting on both such divided regions. 
\erem 

\begin{figure}[htbp]  
\begin{center}
\hspace{-3mm}
\includegraphics[width=0.7\hsize]{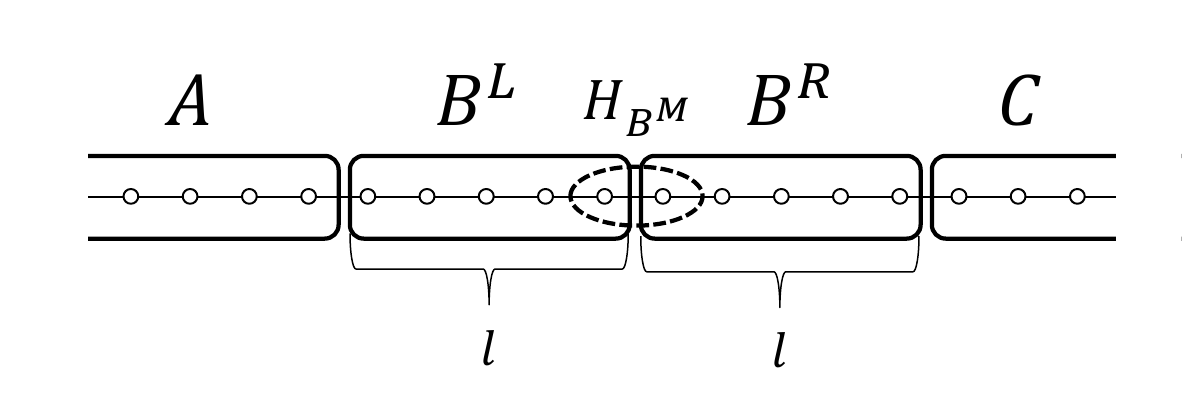}
\vspace{-5mm}
\end{center}
\caption{A schematic picture of the definition of $H_{B^M}$. We divide $B$ into two halves $B^L$ and $B^R$. In the case of a nearest-neighbor Hamiltonian, $H_{B^M}$ is the interaction term acting on both $B^L$ and $B^R$ (the orange circle). }
\label{bh}
\end{figure}

We apply the technical tools discussed in the previous section to a parametrized Hamiltonian $H_{ABC}(s)=H_{AB^L}+H_{B^RC}+sH_{B^M}$.  Here, $H_{B^M}$ corresponds to the perturbation operator $V$ in the previous section. 
Then, we introduce an operator $O_{ABC}$ defined as
\begin{equation*}
O_{ABC}:={\cal T}{\rm Exp}\left[-\frac{\beta}{2}\int_0^1ds\rq{}\Phi_\beta^{H_{ABC}(s\rq{})}\left(H_{B^M}\right)\right]\,.
\end{equation*}
Eq.~\eqref{eq:int} implies that
\begin{equation*}
e^{-\beta H_{ABC}}=O_{ABC}e^{-\beta(H_{AB^L}+H_{B^RC})}O_{ABC}^\dagger\,.
\end{equation*}
We also introduce the inverse operator $\tO_{ABC}$, that is given by 
\begin{equation*}
\tO_{ABC}:={\bar{\cal T}}{\rm Exp}\left[\frac{\beta}{2}\int_0^1ds\rq{}\Phi_\beta^{H_{ABC}(s\rq{})}\left(H_{B^M}\right)\right]\,,
\end{equation*}
where ${\bar {\cal T}}$ denotes the inverse time-ordering operator. 
$O_{ABC}$ and $\tO_{ABC}$ satisfies the following relation (see e.g., \cite{Araki1973})
\begin{equation*}
O_{ABC}\tO_{ABC}=\tO_{ABC}O_{ABC}=\idty_{ABC}\,.
\end{equation*}
From Eq.~\eqref{boundO}, the operator norms of $O_{ABC}$ and $\tO_{ABC}$ can be bounded as
\begin{equation*}
\|O_{ABC}\|, \;\|\tO_{ABC}\|\leq e^{\frac{\beta}{2}\|H_{B^M}\|}=e^{\frac{\beta}{2}J}. 
\end{equation*} 
Importantly, the upper bound is independent of size of $A$, $B$ and $C$. 
From the definitions of $O_{ABC}$ and its inverse, it is not difficult to see that
\begin{equation}\label{idty1}
\rho^{H_{ABC}}=O_{ABC}\left[\tr_{B^RC}\left(\tO_{ABC}\rho^{H_{ABC}}\tO_{ABC}^\dagger\right)\ot\rho_{B^RC}^{H_{B^RC}}\right]O_{ABC}^\dagger\,.
\end{equation}

From Eq.~\eqref{eq:defOs}, we know that there exist operators $O_B$ and $\tO_B$ whose supports are restricted on $B$. 
For simplicity, let us denote
\begin{equation*}
K(\beta)=\frac{c\rq{}\beta J}{2}e^{\frac{(1+c\rq{})\beta J}{2}}\,.
\end{equation*}
Then, $O_B$ and $\tO_B$ satisfy
\begin{equation}\label{approX}
\|O_{ABC}-O_B\|, \|\tO_{ABC}-\tO_B\|\leq K(\beta)e^{-q_1(\beta)l}\,,
\end{equation}
where $q_1(\beta)=\frac{c\rq{}}{1+c\rq{}v\beta/\pi}$ for non-negative constants $c\rq{}$ and $v$ which are chosen as in Eq.~\eqref{eq:defOs}. 
 Then, the operator norm of the local operators $O_B$ and $\tO_B$ can be bounded by
\begin{align}\label{upbob}
\|O_B\|,\|\tO_B\|&\leq \|O_{ABC}\|+\|O_{ABC}-O_B\|\leq e^{\frac{\beta}{2}J}+K(\beta)\,, 
\end{align}
which is independent of the size of $B$. 
Let ${\tilde O}_{B|B}$ be the non-trivial part of ${\tilde O}_B$ acting on $B$, i.e., 
\begin{equation*}
{\tilde O}_{B}={\tilde O}_{B|B}\ot\idty_{AC}\,.
\end{equation*}
By using this notation, we define a CP-map $\kappa_{B\to BC}$ by replacing $O_{ABC}(\tO_{ABC})$ by local operators $O_B(\tO_B)$ and removing partial trace over $C$ in Eq.~\eqref{idty1}, i.e.,
\begin{align}\label{approxcp}
\kappa_{B\to BC}(\sigma_B)\equiv O_{B}\left[\tr_{B^R}\left(\tO_{B|B}\sigma_B\tO_{B|B}^\dagger\right)\ot\rho_{B^RC}^{H_{B^RC}}\right]O^\dagger_{B}\,.
\end{align}

Let us denote 
\begin{equation*}
X_1 = \tr_{B^RC}\left(\tO_{ABC}\rho^{H_{ABC}}\tO_{ABC}^\dagger\right)\ot\rho_{B^RC}^{H_{B^RC}},
\end{equation*}
and
\begin{equation*}
X_2 = \tr_{B^R}\left(\tO_{B|B}\rho^{H_{ABC}}_{AB}\tO_{B|B}^\dagger\right)\ot\rho_{B^RC}^{H_{B^RC}}.
\end{equation*}
We have 
\begin{eqnarray}
\|X_1-X_2\|_1&=& \left\|\tr_{B^RC}\left(\tO_{ABC}\rho^{H_{ABC}}\tO_{ABC}^\dagger\right)\ot\rho_{B^RC}^{H_{B^RC}}\right.\nonumber\\
&&\hspace{3.26cm}\left.-\tr_{B^R}\left(\tO_{B|B}\rho^{H_{ABC}}_{AB}\tO_{B|B}^\dagger\right)\ot\rho_{B^RC}^{H_{B^RC}}\right\|_1
\nonumber\\
&=& \left\|\tr_{B^RC}\left(\tO_{ABC}\rho^{H_{ABC}}\tO_{ABC}^\dagger\right)-\tr_{B^R}\left(\tO_{B|B}\rho^{H_{ABC}}_{AB}\tO_{B|B}^\dagger\right)\right\|_1\nonumber
\\
&=&\left\|\tr_{B^RC}\left(\tO_{ABC}\rho^{H_{ABC}}\tO_{ABC}^\dagger\right)-\tr_{B^RC}\left(\tO_{B}\rho^{H_{ABC}}\tO_{B}^\dagger\right)\right\|_1\nonumber
\\
&\leq&
\left\|\tO_{ABC}\rho^{H_{ABC}}\tO_{ABC}^\dagger-\tO_{B}\rho^{H_{ABC}}\tO_{B}^\dagger\right\|_1\nonumber\\
&\leq&\left\|(\tO_{ABC}-\tO_B)\rho^{H_{ABC}}\tO_{ABC}^\dagger\right\|_1\hspace{-2mm}+\left\|\tO_{B}\rho^{H_{ABC}}(\tO_{ABC}^\dagger-\tO_{B}^\dagger)\right\|_1\label{mideqg}\,.
\end{eqnarray}
We used the monotonicity of the trace-norm in the last inequality. 
To address the calculation, we use the following spacial case of the H$\ddot{\rm o}$lder\rq{}s inequality:
\begin{equation}\label{Holder}
\|AB\|_1\leq\|A\|_1\|B\|\,.
\end{equation}
It implies that
\begin{eqnarray*}
\hspace{-6mm}\eqref{mideqg}&\leq&\left\|(\tO_{ABC}-\tO_B)\right\|\left\|\rho^{H_{ABC}}\tO_{ABC}^\dagger\right\|_1+\left\|\tO_{B}\rho^{H_{ABC}}\right\|_1\left\|(\tO_{ABC}^\dagger-\tO_{B}^\dagger)\right\|\\
&\leq&\left\|(\tO_{ABC}-\tO_B)\right\|\left\|\tO_{ABC}^\dagger\right\|+\left\|\tO_{B}\right\|\left\|(\tO_{ABC}^\dagger-\tO_{B}^\dagger)\right\|\\
&\leq& 2K(\beta)\left(e^\frac{\beta J}{2}+K(\beta)\right)e^{-q_1(\beta)l}\,.
\end{eqnarray*}
The first and second lines follow from Eq.~\eqref{Holder} and $\|\rho^{H_{ABC}}\|_1=1$. 
In the last line, we used Eq.~\eqref{approX} and Eq.~\eqref{upbob}.

By using the above bound, we bound the difference between the original Gibbs state $\rho^{H_{ABC}}$ and $\kappa_{B\to BC}(\rho^{H_{ABC}}_{AB})$ as
\begin{eqnarray*}
&& \left\|\rho^{H_{ABC}}-\kappa_{B\to BC}(\rho_{AB}^{H_{ABC}})\right\|_1=\left\|O_{ABC}X_1O_{ABC}^\dagger-O_BX_2O_B^\dagger\right\|_1\\
&\leq&\left\|O_{ABC}X_1O_{ABC}^\dagger-O_BX_1O_B^\dagger\right\|_1+\left\|O_{B}(X_1-X_2)O_B^\dagger\right\|_1\\
&\leq&\left\|(O_{ABC}-O_B)X_1O_{ABC}^\dagger\right\|_1+\left\|O_BX_1(O_{ABC}^\dagger-O^\dagger_B)\right\|_1+\left\|(X_1-X_2)\right\|_1\left\|O_B^\dagger\right\|^2\\
&\leq&\|O_{ABC}-O_B\|\|{\tilde O}_{ABC}\|^2\|O_{ABC}\|+\|O_{ABC}^\dagger-O^\dagger_B\|\|{\tilde O}_{ABC}\|^2\|O_{B}\|\nonumber\\
&\quad&+\left\|(X_1-X_2)\right\|_1\left\|O_B^\dagger\right\|^2\\
&\leq& K(\beta)\left(e^{\frac{3\beta J}{2}}+e^{\beta J}\left(e^{\frac{\beta J}{2}}+K(\beta)\right)+2\left(e^{\frac{\beta J}{2}}+K(\beta)\right)^3\right)e^{-q_1(\beta)l}\\
&\leq&  4K(\beta)\left(e^{\frac{\beta J}{2}}+K(\beta)\right)^3e^{-q_1(\beta)l}\,.
\end{eqnarray*}
Here we used the fact that $\|X_1\|_1\leq\|\rho^{H_{ABC}}\|_1\|\tO_{ABC}\|^2=\|\tO_{ABC}\|^2$ in the fourth line. Choosing $C_1(\beta)=4K(\beta)\left(e^{\frac{\beta J}{2}}+K(\beta)\right)^3$ completes the proof.
\end{proof}

Unfortunately, the map $\kappa_{B\to BC}$ is not a trace non-increasing map in general. 
Thus, we normalize $\kappa_{B\to BC}$ to obtain a CP and trace non-increasing map, which corresponds to a probabilistic process. 
\blm\label{l.lemma2}
Under the setting of lemma~\ref{lem1}, there exists a CP and trace non-increasing map $\tLambda_{B\to BC}$ for any $l\geq l_0(\beta)\equiv\left\lceil\frac{\log C_1(\beta)+1}{q_1(\beta)}\right\rceil=\cO(\beta^2)$ such that
\begin{equation}\label{lemma2}
\left\|\rho^{H_{ABC}}-\frac{\tLambda_{B\to BC}(\rho^{H_{ABC}}_{AB})}{\tr[\tLambda_{B\to BC}(\rho^{H_{ABC}}_{AB})]}\right\|_1 \leq C_2(\beta)e^{-q_1(\beta)l}\,,
\end{equation}
where $C_2(\beta)=\frac{2C_1(\beta)}{(1-e^{-1})}$. 
Moreover, $p=\tr[\tLambda_{B\to BC}(\rho^{H_{ABC}}_{AB})]$ is a strictly positive constant which is independent of the size of subsystems $A$, $B$ and $C$.
\elm
\begin{proof}
We denote the maximum eigenvalue of $O_B^\dagger O_B$ ($\tO_B^\dagger \tO_B$) by $\lambda_{\max}^{O_B}$ ($\lambda_{\max}^{\tO_B}$). 
From Eq.~\eqref{upbob}and inequality $\|A^\dagger A\|\leq\|A\|^2$, these eigenvalues are bounded as 
\begin{equation}\label{boundlamo}
\lambda_{\max}^{O_B}, \lambda_{\max}^{\tO_B}\leq\left(e^{\frac{\beta J}{2}}+K(\beta)\right)^2\,.
\end{equation}
Define $\lambda_{\max}:=\lambda_{\max}^{O_B}\lambda_{\max}^{\tO_B}$. Then, we define the normalized map $\tLambda_{B\to BC}$ as
\begin{equation*}
\tLambda_{B\to BC}(\sigma_B):=\frac{1}{\lambda_{\max}}\kappa_{B \to BC}(\sigma_B)\,.
\end{equation*}
By definition, $\tLambda_{B\to BC}$ is CP and trace non-increasing. 
After the normalization, the output state for the input $\rho^{H_{ABC}}_{AB}$ is 
\begin{equation*}
\frac{\tLambda_{B\to BC}(\rho^{H_{ABC}}_{AB})}{\tr[\tLambda_{B\to BC}(\rho^{H_{ABC}}_{AB})]}\,.
\end{equation*}

Let us introduce $l_0(\beta)=\frac{\ln C_1(\beta)+1}{q_1(\beta)}$. For any $l\geq l_0(\beta)$, $C_1(\beta)e^{-q_1(\beta)l}\leq e^{-1}<1$. For such $l$, the probability $p$ for the input $\rho^{H_{ABC}}_{AB}$ is then estimated as
\begin{align}
p&=\tr[\tLambda_{B\to BC}(\rho^{H_{ABC}}_{AB})]\nonumber\\
&=\frac{1}{\lambda_{\max}}\left\|\kappa_{B\to BC}(\rho^{H_{ABC}}_{AB})\right\|_1\nonumber\\
&\geq\frac{1}{\lambda_{\max}}\left|\|\rho^{H_{ABC}}\|_1-\left\|\rho^{H_{ABC}}-\kappa_{B\to BC}(\rho^{H_{ABC}}_{AB})\right\|_1\right|\nonumber\\
&\geq\frac{1}{\lambda_{\max}}\left(1-C_1(\beta)e^{-q_1(\beta)l}\right)\nonumber\\
&\geq\frac{1}{\lambda_{\max}}\left(1-\frac{1}{e}\right)\nonumber\\
&\geq\frac{1-e^{-1}}{\left(e^{\frac{\beta J}{2}}+K(\beta)\right)^4}\label{Pbound}\\
&>0\nonumber\,,
\end{align}
where we used Eq.~\eqref{boundlamo} in the line before the last. 

The approximation error of the output is then  estimated as
\begin{align*}
&\left\|\rho^{H_{ABC}}-\frac{\tLambda_{B\to BC}(\rho^{H_{ABC}}_{AB})}{\tr[\tLambda_{B\to BC}(\rho^{H_{ABC}}_{AB})]}\right\|_1
=\left\|\rho^{H_{ABC}}-\frac{\kappa_{B\to BC}(\rho^{H_{ABC}}_{AB})}{\|\kappa_{B \to BC}(\rho^{H_{ABC}}_{AB})\|_1}\right\|_1\\
&\leq\left|1-\frac{1}{\|\kappa_{B \to BC}(\rho^{H_{ABC}}_{AB})\|_1}\right|\|\rho^{H_{ABC}}\|_1\nonumber\\
&\hspace{3cm}+\frac{1}{\|\kappa_{B \to BC}(\rho^{H_{ABC}}_{AB})\|_1}\left\|\rho^{H_{ABC}}-\kappa_{B\to BC}(\rho^{H_{ABC}}_{AB})\right\|_1\\
&\leq\frac{C_1(\beta)e^{-q_1(\beta)l}}{1-e^{-1}}+\frac{1}{1-e^{-1}}C_1(\beta)e^{-q_1(\beta)l}\\
&\leq\frac{2C_1(\beta)}{1-e^{-1}}e^{-q_1(\beta)l}\,.
\end{align*}
In the third line,  we used the fact 
\begin{equation*}
\left|\|\kappa_{B\to BC}(\rho_{AB}^{H_{ABC}})\|_1-1\right|\leq C_1(\beta)e^{-q_1(\beta)l}\,,
\end{equation*}
which follows from
\begin{equation*}
1-C_1(\beta)e^{-q_1(\beta)l}\leq\|\kappa_{B\to BC}(\rho_{AB}^{H_{ABC}})\|_1\leq1+C_1(\beta)e^{-q_1(\beta)l}\,.
\end{equation*}
Thus, we conclude that Lemma~\ref{l.lemma2} holds by choosing $C_2(\beta)=\frac{2C_1(\beta)}{1-e^{-1}}$.
\end{proof}

\begin{figure}[htbp]  
\begin{center}
\includegraphics[width=0.9\textwidth]{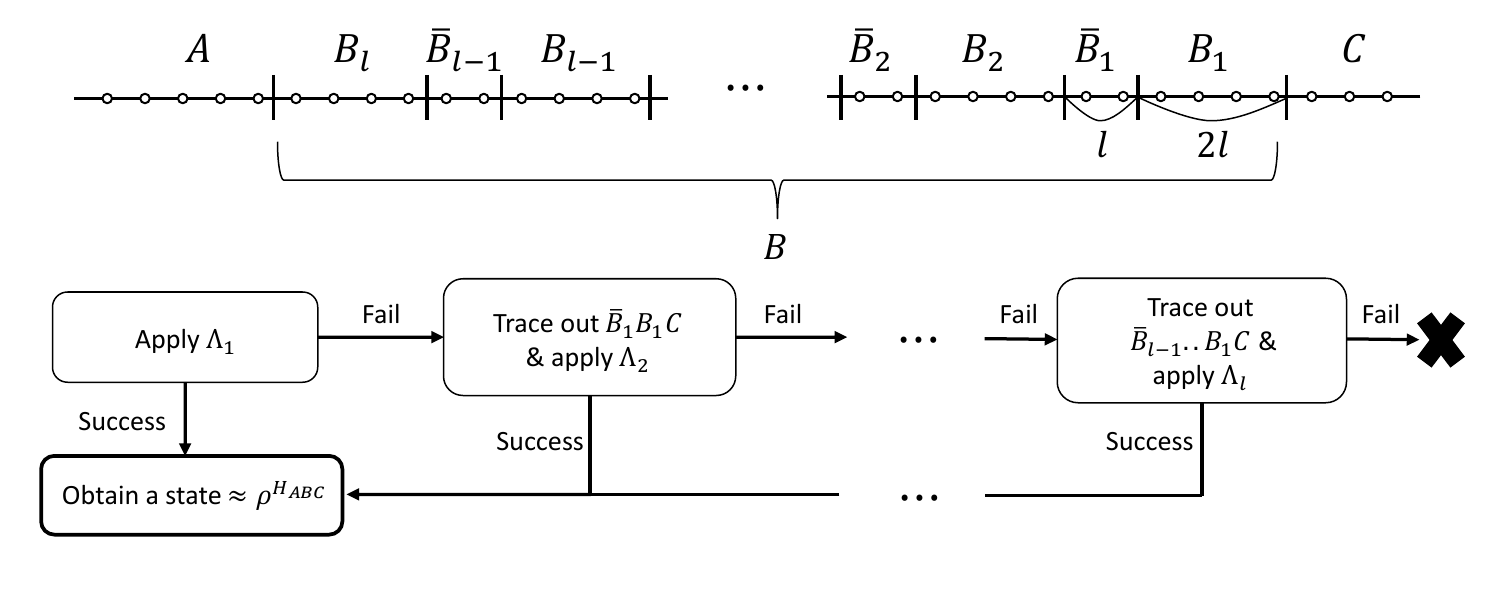}
\vspace{-10mm}
\end{center}
\caption{A schematic picture of the repeat-until-success method. We introduced buffer systems $\{{\bar B}_i\}$ to suppress the effect of failure. The \lq\lq{}failure\rq\rq{} output at the end corresponds to the CP-map $\tE_l\circ\tr_{l-1}\tE_{l-1}\cdots\circ\tr_1\tE_1$.}
\label{rus}
\end{figure}

We are now in position to prove Theorem~\ref{mthm1}. 
Without loss of generality, let us assume that $d(A,C)=|B|=3l^2-l$ for $l\in{\mathbb N}$. 
We divide $B$ into $B=B_l{\bar B}_{l-1}B_{l-1}...{\bar B}_1B_1$ as shown in Fig.~\ref{rus},  where for each $i$, $|B_i|=2l$ and $|{\bar B_i}|=l$. From Lemma~\ref{l.lemma2},  there exists a CP and trace non-increasing map $\tLambda_{B_i\to B_i{\bar B_{i-1}}...B_1C}$ for each $i$ which approximately recovers $\rho^{H_{ABC}}$ from the reduced state on $AB_l...{\bar B}_iB_i$ (here, we choose $B_i$ as ``$B$'' and $B_{i-1}...B_1C$ as ``$C$'' in the lemma). 
We then introduce a CP and trace non-increasing map $\tE_{B_i\to B_i{\bar B_{i-1}}...B_1C}$ for each $\tLambda_{B_i\to B_i{\bar B_{i-1}}...B_1C}$ so that $\{\tLambda_{B_i\to B_i{\bar B_{i-1}}...B_1C}, \tE_{B_i\to B_i{\bar B_{i-1}}...B_1C}\}$ forms a quantum  instrument~\footnote{For instance, $\tE_{B \to BC}$ can be chosen to be $\tE_{B \to BC}(\sigma_B)=\sqrt{\idty-M_B}\sigma_B\sqrt{\idty-M_B}\ot|0\rangle\langle0|_C$, where $M_B:=\sum_iK_i^\dagger K_i\leq \idty_B$ with $\{K_i\}$ the Kraus operators of  $\tLambda_{B\to BC}$. }.  
We simply denote $\tLambda_{B_i\to B_i{\bar B_{i-1}}...B_1C}$ by $\tLambda_i$ and its complementary map $\tE_{B_i\to B_i{\bar B_{i-1}}...B_1C}$ by $\tE_i$. We also denote the tracing operation over ${\bar B_i}B_i...B_1C$ by $\tr_i$. 
We define a CPTP-map $\Lambda_{B\to BC}$ as 
\begin{align}
&\Lambda_{B\to BC}(\sigma_{B})=\nonumber\\
&\tLambda_1(\sigma_B)+\left(\tLambda_2+\cdots\left(\tLambda_{l-1}+\left(\tLambda_{l}+\tE_l\right)\tr_{l-1}\tE_{l-1}\right)\cdots\tr_2\tE_2\right)\tr_1\tE_1(\sigma_B)\,\label{rusmap}
\end{align}
based on the repeat-until-success method (Fig.~\ref{rus}). 

When we input $\rho^{H_{ABC}}_{AB}$ to $\Lambda_{B\to BC}$, the output of each map $\tLambda_i$ corresponds to the success of the recovery process at the $i$th step (with probability $p_i$) and $\tE_i$ corresponds to the failure of the recovery process (with probability $1-p_i$). If it fails, we trace out both the recovered system and the \lq\lq{}buffer\rq\rq{} system ${\bar B}_i$, and then, the effect of the failure can be almost neglected. 
Thus, we can repeat the probabilistic process to obtain the success outcome. 
The effect of the failure is estimated by the following lemma, which utilizes the exponential decay of correlation of 1D Gibbs states~\cite{Araki1969}. 

\blm\label{ch3le3}
Under the setting of Lemma~\ref{lem1}, there exists a constant $\xi\geq0$ such that
\begin{equation*}
(1-p_i)\left\|\tr_i\left(\rho^{H_{ABC}}\right)-\frac{\tr_i\tE_i\left(\rho^{H_{ABC}}_{AB_l...B_i}\right)}{1-p_i}\right\|_1\leq e^{-\frac{l}{\xi}}\,.
\end{equation*}
\elm
\begin{proof}
Define a correlation function ${\rm Cor}(X:Y)_\rho$ of regions $X$ and $Y$ by
\begin{equation*}
{\rm Cor}(X:Y)_\rho=\max_{\|M\|,\|N\|\leq1}\left|\tr\left[(M\ot N)(\rho_{XY}-\rho_X\ot \rho_Y)\right]\right|\,.
\end{equation*}
Consider some CP and trace-decreasing map $(\idty_X\ot{\cal E}_{Y\to Z})(\rho_{XY})=\sum_iE_i\rho_{XY}E_i^\dagger$. By lemma 9 of Ref.~\cite{Brandao2015}, it holds that
\begin{equation*}
{\rm Cor}(X:Y)_\rho\geq \tr[M_Y\rho_{XY}]\|\rho_X-\sigma_X\|_1\,,
\end{equation*}
where $M_Y=\sum_iE_i^\dagger E_i$ and 
\begin{equation*}
\sigma_X=\frac{1}{\tr[M_Y\rho_{XY}]}\tr_Y\left[M_Y\rho_{XY}\right]=\tr_Y\frac{(\id_X\ot{\cal E}_{Y\to Z})(\rho_{XY})}{\tr(\id_X\ot{\cal E}_{Y\to Z})(\rho_{XY})}\,.
\end{equation*}
It has been shown that any 1D Gibbs states with a short-range Hamiltonian have exponentially decaying ${\rm Cor}(X:Y)_\rho$~\cite{Araki1969}, i.e., there exist constants $c,\xi\geq0$ such that
\begin{equation}\label{edccf}
{\rm Cor}(X:Y)_\rho\leq ce^{-d(X:Y)/\xi}\,.
\end{equation} 
Choosing $X=AB_l...B_{i+1}$, $Y=B_i$, $Z=B_i...B_1C$ and ${\cal E}_{Y\to Z}=\tE_i$ prove Lemma~\ref{ch3le3}. 
\end{proof}

Without loss of generality, let us assume 
\begin{align*}
p_i=\tr[\tLambda_i(\rho^{H_{ABC}}_{AB_l...B_i})]=p>0
\end{align*}
for all $i$~\footnote{Ohterwise, we just pick the smallest $p$ among $p_i,i=1,...,l$ and redefine $\tLambda_i$ to be $\frac{p}{p_i}\tLambda_i$ and $\tE_i$ to be $\tE_i+(1-\frac{p}{p_i})\tLambda_i$.}. 
Lemma~\ref{ch3le3} allows an iterative calculation. First we have
\begin{align*}
&\left\|\Lambda_{B\to BC}(\rho^{H_{ABC}}_{AB})-\tLambda_1(\rho^{H_{ABC}}_{AB})+(1-p)\tLambda_2(\rho^{H_{ABC}}_{AB_l...B_2})\right.\nonumber\\
&\hspace{1cm}\left.+(1-p)\left(\tLambda_3+\left(\cdots\left(\tLambda_{l}+\tE_l\right)\tr_{l-1}\tE_{l-1}\right)\cdots\right)\tr_2\tE_2(\rho^{H_{ABC}}_{AB_l...B_2})\right\|_1\leq e^{-l/\xi}\, 
\end{align*}
Here, we used Lemma~\ref{ch3le3} for $\tr_1\tE_1(\rho_{AB}^{H_{ABC}})$ and $\rho_{AB_l...B_2}^{H_{ABC}}$. 
Then we can obtain
\begin{align*}
&\left\|\Lambda_{B\to BC}(\rho^{H_{ABC}}_{AB})-\tLambda_1(\rho^{H_{ABC}}_{AB})+(1-p)\tLambda_2(\rho^{H_{ABC}}_{AB_l...B_2})\right.\nonumber\\
&\left.\hspace{1cm}+(1-p)^2\tLambda_3(\rho^{H_{ABC}}_{AB_l...B_3})+(1-p)^2\left(\tLambda_4+\cdots\tr_4\tE_4\right)\tr_3\tE_3(\rho^{H_{ABC}}_{AB_l...B_3})\right\|_1\leq 2e^{-l/\xi}\,. 
\end{align*}
We can proceed in a similar way, where at each $i$th step, we replace $\tr_i\tE_i(\rho_{AB}^{H_{ABC}})$ by $\tr_i(\rho^{H_{ABC}})$ by using the triangle inequality. After iterating $l-1$ steps, we obtain
\begin{align}
&\left\|\Lambda_{B\to BC}(\rho^{H_{ABC}}_{AB})-\sum_{i=1}^l(1-p)^{i-1}\tLambda_i(\rho^{H_{ABC}}_{AB})+(1-p)^{l-1}\tE_l(\rho^{H_{ABC}}_{AB_l})\right\|_1\leq (l-1)e^{-l/\xi}\,.\label{eq:f1}
\end{align}
Since $\left(\sum_{i=1}^{l}p(1-p)^{i-1}\right)+(1-p)^l=1$, it follows 
\begin{equation*}
\rho^{H_{ABC}}=\sum_{i=1}^{l}p(1-p)^{i-1}\rho^{H_{ABC}}+(1-p)^l\rho^{H_{ABC}} 
\end{equation*}
and thus
\begin{align}
&\left\|\rho^{H_{ABC}}-\sum_{i=1}^l(1-p)^{i-1}\tLambda_i(\rho^{H_{ABC}}_{AB})+(1-p)^{l-1}\tE_l(\rho^{H_{ABC}}_{AB_l})\right\|_1\nonumber \\
&\leq\sum_{i=1}^lp(1-p)^{i-1}\left\|\rho^{H_{ABC}}-\frac{1}{p}\tLambda_i(\rho^{H_{ABC}}_{AB_l...B_i})\right\|_1 +(1-p)^l\left\|\rho^{H_{ABC}}-\frac{1}{1-p}\tE_l(\rho^{H_{ABC}}_{AB_l})\right\|_1 \nonumber\\
&\leq \{1-(1-p)^{l}\}C_2(\beta)e^{-q_1(\beta)l}+2(1-p)^l\,.\label{eq:f2}
\end{align}
Therefore, by combining Eq.~\eqref{eq:f1} and Eq.~\eqref{eq:f2}, we conclude 
\begin{align}
\|\rho^{H_{ABC}}-\Lambda_{B\to BC}(\rho^{H_{ABC}}_{AB})\|_1 &\leq \{1-(1-p)^l\}C_2(\beta)e^{-q_1(\beta)l}+2e^{-|\ln(1-p)|l}+(l-1)e^{-l/\xi}\nonumber\\
&\leq C_2(\beta)e^{-q_1(\beta)l}+2e^{-|\ln(1-p)|l}+le^{-l/\xi}\,.\label{eq:f3}
\end{align}
Here, the probability $p$ can be bounded as in Eq.~\eqref{Pbound}, and thus we have
\begin{equation*}
|\ln(1-p)|\geq\left|\ln\left(1-\frac{(1-e^{-1})e^{-2\beta J}}{\left(1+e^{-\frac{\beta J}{2}}K(\beta)\right)^4}\right)\right|\geq \frac{(1-e^{-1})e^{-2\beta J}}{\left(1+e^{-\frac{\beta J}{2}}K(\beta)\right)^4}=e^{-\Theta(\beta)}\,,
\end{equation*}
 where the last inequality follows from $\log(1-x)\leq -x$ for any $x\in[0,1]$. 
If $\xi= e^{\cO(\beta)}$,  Eq.~\eqref{eq:f3} can be bounded by
\begin{equation}\label{upboundf}
2C_2(\beta)le^{-q\rq{}(\beta)l}=e^{-q\rq{}(\beta)l+\ln(2C_2(\beta)l)}\,,
\end{equation}
where $q\rq{}(\beta)=\Omega( e^{-\Theta(\beta)})$. 
Since $d(A,C)=3l^2-l$, Eq.~\eqref{upboundf} is $\Theta(e^{-\Theta(\sqrt{d(A,C)})})$. Therefore, for sufficiently large $l$, there exists a constant $q(\beta)=\Omega( e^{-\Theta(\beta)})$ such that $e^{-q\rq{}(\beta)l+\ln(2C_2(\beta)l)}\leq e^{-q(\beta)\sqrt{d(A,C)}}$.

\subsection{Proof of Corollary \ref{cor1}}

\begin{proof}
Let us first consider a 1D open spin chain with a tripartition $ABC$ so that a simply connected region $B$ shields $A$ from $C$. Then, $d(A,C)=|B|$. Without loss of generality, we assume $|A|\leq |B|\leq |C|$. 
Divide $C$ into $C=C_1\cup C_2\cup...\cup C_m$, where $m$ is the maximum number such that $|C_i|=|B|$ for $1\leq i< m$ and each $C_i$ shields $C_{i-1}$ from $C_{i+1}$ (here, $C_0\equiv B$). 

Theorem \ref{m.mthm2} and the Fannes inequality imply 
\begin{align*}
I(A : C_i | B C_1 \ldots C_{i-1})_{\rho^{H_{ABC}}} &= S(A|B C_1 \ldots C_{i-1})_{\rho^{H_{ABC}}}-S(A|BC_1...C_i)_{\rho^{H_{ABC}}}\\
&\leq S(A|B C_1 \ldots C_{i})_{\Lambda_{B...C_{i-1}\to BC}(\rho_{AB...C_{i-1}}^{H_{ABC}})}-S(A|BC_1...C_i)_{\rho^{H_{ABC}}}\\
&\leq 6|B| e^{ -\frac{q(\beta) }{2}\sqrt{i|B|}} 
\end{align*}
with a constant $q(\beta)\geq0$ for any $i\in[1, m]$ and sufficiently large $|B|$. 
By the chain rule 
\begin{equation*}
I(A:C|B)  = I(A:C_1|B)  + I(A:C_2|BC_1)  + \ldots +  I(A:C_m | B C_1 \ldots C_{m-1})  \,,
\end{equation*}
we have 
\begin{eqnarray*}
I(A:C|B)_{\rho^{H_{ABC}}}&\leq& 6\sum_{i=1}^m |B| e^{-\frac{q(\beta) }{2} \sqrt{i|B|}}  \nonumber  \\
&\leq& 6\left( |B| e^{-\frac{q(\beta) }{2}\sqrt{|B|}} + |B|\sum_{i=1}^{m-1} e^{-\frac{q(\beta) }{2} \sqrt{(i+1)|B|}}\right) \nonumber \\
&\leq& 6\left( |B|e^{-\frac{q(\beta) }{2}\sqrt{|B|}}+\int_1^\infty e^{-\frac{q(\beta) }{2}\sqrt{x|B|}}dx \right)   \nonumber \\
&=& 6\left(|B|+\frac{8(1+\frac{q(\beta) }{2}\sqrt{|B|})}{q(\beta)^2}\right) e^{-\frac{q(\beta) }{2}\sqrt{|B|}}.
\end{eqnarray*}
Again, the upper bound is $e^{-\Theta(\sqrt{d(A,C)})}$. The same strategy works for a more complicated tripartition $ABC$ of both 1D open chains and closed chains.
\end{proof}

\subsection{Proof of Corollary \ref{lowdepth preparation}}\label{lowdeproof}

\begin{proof}Let us consider 1D open spin chain without loss of generality. 
We first divide the whole chain into consecutive regions $A_1B_1C_1A_2B_2C_2...A_kB_kC_k$, where we choose $|A_i|=|B_j|=l\geq l_0$ and $|C_i|=5\xi(\ln d) l$ for all $i$, where 
$l_0$ is the constant given in Theorem~\ref{m.mthm2}, the correlation length $\xi$ is given in Eq.~\eqref{edccf} and $d$ is a constant bounding the dimension of the Hilbert spce of one spin from above. 
Let us consider region $(A_1B_1C_1...C_{i-1}A_iB_{i+1})(B_iA_{i+1})C_i$, where $B_iA_{i+1}$ shields $A_1B_1...B_{i+1}$ from $C_i$. 
From Theorem~\ref{m.mthm2}, there exists a CPTP-map $\delta_i:\cD(\cH_{B_iA_{i+1}})\to\cD(\cH_{B_iC_iA_{i+1}})$ such that 
\begin{align}\label{FRapplication}
\left\|\Delta_i\left(\rho^H_{A_1B_1...A_iB_iA_{i+1}B_{i+1}C_{i+1}}\right)-\rho^H_{A_1B_1...A_iB_iC_iA_{i+1}B_{i+1}C_{i+1}}\right\|_1\leq Ce^{-q(\beta)\sqrt{l}}\,.
\end{align}

Since the Gibbs state has exponentially decaying correlations, after tracing out $C_i$, the two remained connected components are almost uncorrelated.  
By using Lemma 20 of Ref.~\cite{Brandao2015}, it follows that
\begin{align}
&\left\|\rho_{A_1B_1C_1...B_{i-1}A_iB_i}-\rho_{A_1B_1C_1...B_{i-1}}\ot\rho_{A_iB_i}\right\|_1\nonumber\\
&\quad\hspace{0cm}\leq\left(\dim\cH_{A_iB_i}\right)^2Cor(A_1B_1C_1...B_{i-1}:A_iB_i)_{\rho^H}\nonumber\\
&\quad\hspace{0cm}\leq d^{4l}e^{-5\xi(\ln d) l/\xi}\nonumber\\
&\quad\hspace{0cm}=e^{-(\ln d) l}\,.\label{decaycorrelations}
\end{align}

Each $\Delta_i$ acts on different sets of spins and therefore does not overlap. 
Then we have 
\begin{align*}
&\left\|   \Delta_1 \otimes \cdots \otimes \Delta_k( \rho_{A_1B_1} \otimes \cdots \otimes \rho_{ A_k B_k}) - \rho_{A_1B_1C_1 \cdots A_k B_k C_k} \right\|_1  \nonumber\\ 
&\leq \left\| \Delta_1 \otimes \cdots \otimes \Delta_k (\rho_{A_1B_1} \otimes \cdots \otimes \rho_{ A_k B_k}) -\Delta_2 \otimes \cdots \otimes \Delta_k (\rho_{A_1B_1C_1 A_2 B_2} \otimes \rho_{A_3B_3} \otimes \cdots \otimes \rho_{A_kB_k}) \right\|_1 \nonumber \\
&\quad+\left\| \Delta_2 \otimes \cdots \otimes \Delta_k (\rho_{A_1B_1C_1 A_2 B_2} \otimes \rho_{A_3B_3} \otimes \cdots \otimes \rho_{A_kB_k}) -  \rho_{A_1B_1C_1 \ldots A_k B_k C_k}  \right\|_1\nonumber \\
&\leq  \left\|  \Delta_1 (\rho_{A_1B_1} \otimes \rho_{A_2B_2}) - \rho_{A_1B_1C_1 A_2 B_2} \right\|_1 \nonumber \\
&\quad+ \left\| \Delta_2 \otimes \cdots \otimes \Delta_k (\rho_{A_1B_1C_1 A_2 B_2} \otimes \rho_{A_3B_3} \otimes \cdots \otimes \rho_{A_kB_k}) -  \rho_{A_1B_1C_1 \cdots A_k B_k C_k}  \right\|_1  \nonumber\\
&\leq \Vert  \Delta_1 (\rho_{A_1B_1 A_2B_2}) - \rho_{A_1B_1C_1 A_2 B_2} \Vert_1 + e^{-(\ln d)l} \nonumber \\
&\quad+ \Vert \Delta_2 \otimes \cdots \otimes \Delta_k (\rho_{A_1B_1C_1 A_2 B_2} \otimes \rho_{A_3B_3} \otimes \cdots \otimes \rho_{A_kB_k}) -  \rho_{A_1B_1C_1 \cdots A_k B_k C_k}  \Vert_1  \nonumber\\
&\leq 2e^{- q(\beta)\sqrt{l}} +  \Vert \Delta_2 \otimes \cdots \otimes \Delta_k (\rho_{A_1B_1C_1 A_2 B_2} \otimes \cdots \otimes \rho_{A_kB_k}) -  \rho_{A_1B_1C_1 \cdots A_k B_k C_k}  \Vert_1. 
\end{align*}
The first inequality follows from the triangle inequality, the second from the monotonicity of the trace norm under quantum operations, the third from Eq. (\ref{decaycorrelations}), and the fourth from Eq. (\ref{FRapplication}) and $e^{-(\ln d)l}\leq e^{- q(\beta)\sqrt{l}}$ for large $l$.  

Iterating the argument above, we find 
\begin{equation*}
\left\|   \Delta_1 \otimes \cdots \otimes \Delta_k( \rho_{A_1B_1} \otimes \ldots \otimes \rho_{ A_k B_k}) - \rho_{A_1B_1C_1 \ldots A_k B_k C_k} \right\|_1 \leq  2ke^{- q(\beta)\sqrt{l}}\,.
\end{equation*}

Since $k \leq n$, choosing $l = \cO(\log^2(n/\varepsilon))$ gives an error bounded by $\varepsilon$. 
We denote a CPTP-map which construct $\rho_{A_iB_i}$ by $\Delta_{1,i}$ and relabel $\Delta_i$ in the above by $\Delta_{2,i}$. 
The CPTP-map $\bigotimes_i\Delta_{1,i}$ creates product state of the form of $\rho_{A_1B_1}\ot\rho_{A_2B_2}\ot\cdots\ot\rho_{A_kB_k}$, and then $\bigotimes_i\Delta_{2, i}$ approximately creates the target state from this product state. 
\end{proof}

\section{Extension to More General Graphs and A Conjecture for Higher Dimension}\label{sec:etmgg}
\begin{figure}[htbp]  
\begin{center}
\vspace{-3mm}
\includegraphics[width=10cm]{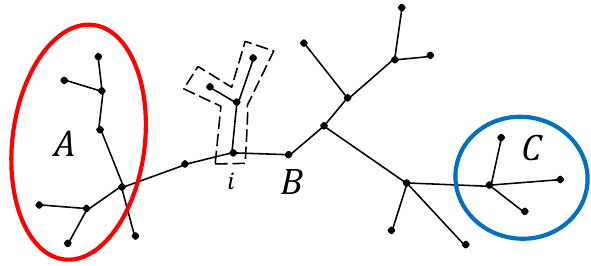}
\vspace{-5mm}
\end{center}
\caption{An example of the region with a partition $ABC$ for a tree graph. Here, $B$ is the set of all spins in outside of the circles. In the coarse-grain procedure, the descendants of $i$ (the spins in the dashed region) can be regarded as one large system.}
\label{tree1}
\end{figure}

Our proof for 1D spin chains can be generalized to more general graphs with appropriate partitions. 
For instance, let us consider a tree graph $G=(E, V)$ with a partition $ABC$ as depicted in Fig.~\ref{tree1}.  
Since $G$ is a tree, there is a unique path connecting $A$ and $C$. Then, all spins in $B$ are classified as $(i)$ spins belonging to the path $(ii)$ descendants of spins on the path $(iii)$ the rest spins which are separated from the path. We can obtain a coarse grained 1D chain by regarding each spin on the path and its descendants as one system, and removing all spins in $(iii)$. Therefore, we can apply the proof in the previous section to this situation as well. Note that the norm of an interaction term connecting spins on the path is irrelevant to the size of the coarse grained spins.

\begin{figure}[htbp]  
\begin{center}
\includegraphics[width=6cm]{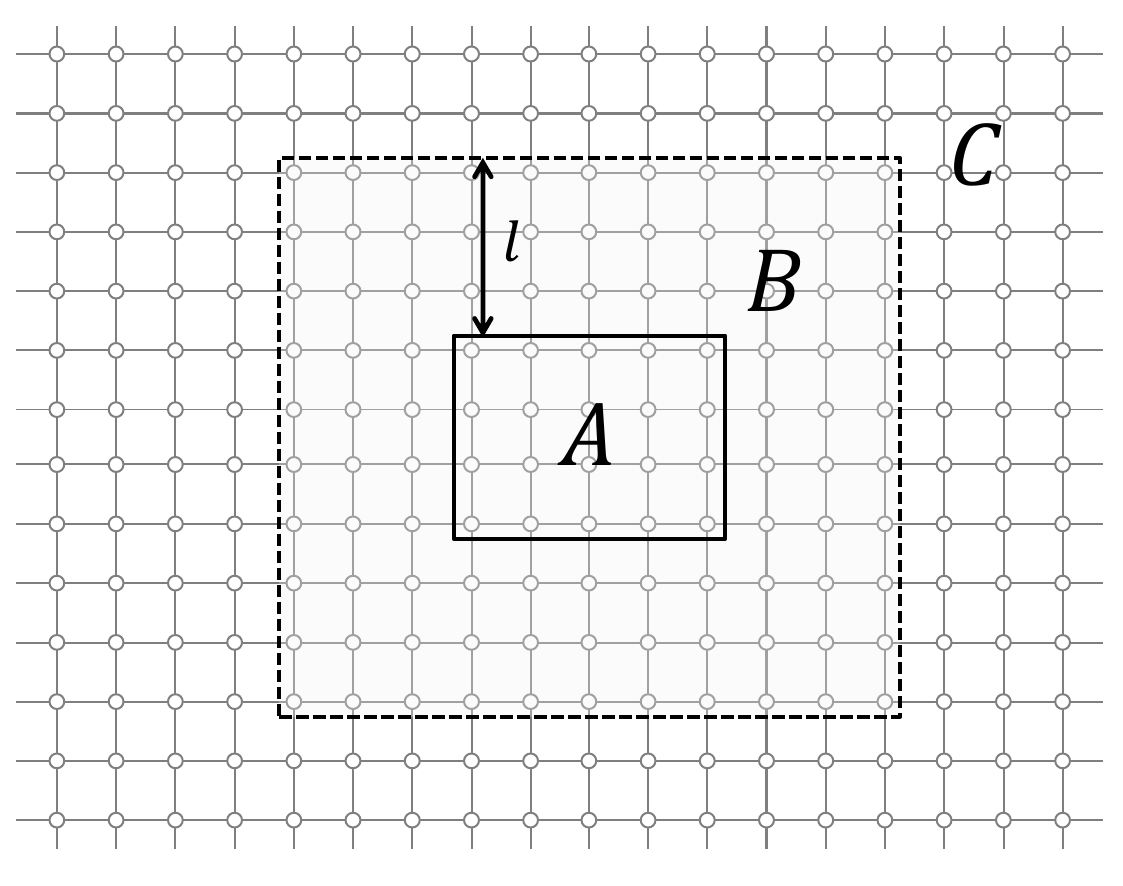}
\vspace{-5mm}
\end{center}
\caption{An example of  a 2D lattice with a partition $ABC$. We expect that the conditional mutual information $I(A:C|B)_\rho$ for any Gibbs state decays fast with respect to $d(A, C)$.}
\label{lattice2d}
\vspace{-5mm}
\end{figure}

An important point of the above argument is that the success probability of the recovery map in lemma~\ref{l.lemma2} is bounded by a constant of $d(A,C)$. 
In general cases, we can consider a partition $ABC$ which cannot be reduced to 1D systems, such as depicted in Fig.~\ref{lattice2d}. 
Remember that the success probability $p$ is in the order $\Omega(e^{\beta\|H_{B^M}\|})$.  
In the case of Fig.~\ref{lattice2d}, $H_{B^M}$ is the sum of all interactions along the perimeter of $B^L$, {which is proportional to $l$}. 
When considering the repeat-until-success method, the success probability decays too rapidly, and therefore our strategy does not work.

The quantum Hammersley-Clifford theorem holds for more general class of Markov networks such as regular $D$-dimensional lattices. 
In this case, a partition $ABC$ of the system is chosen so that $B$ shields $A$ from $C$ (Figure~\ref{lattice2d}).  Due to this observation, we expect that Gibbs states of short-ranged Hamiltonians obey fast decay of the conditional 
mutual information in $D$-dimensional systems as well. We have the following conjecture:

\begin{conjecture}
Let $\rho$ be a Gibbs state of a short-ranged Hamiltonian defined on a $D$-dimensional spin lattice. 
Then, there exist constants $C, c>0$ such that for every three regions $A, B, C$ with $B$ shielding $A$ from $C$,
\begin{equation*}
I(A:C|B)_\rho\leq C e^{-cd(A,C)}\,
\end{equation*}
\end{conjecture}

Note that when $D=1$, this conjecture gives an improved bound. It turns out that the area law for mutual information implies a weak version of the conjecture, as discussed in Sec.~\ref{sec:satuarea}. Consider a Gibbs state in the infinite volume limit (as a KMS state). Let $A$ be a region of the lattice and $B_l$ be a ring around $A$ of width $l$. 
Then, because of the area law~\cite{PhysRevLett.100.070502}, for every $\varepsilon>0$ there is an integer $l$ s.t. 
\begin{equation*}
I(A:B_{l+1}) - I(A:B_l) \leq \varepsilon, 
\end{equation*}
which can be written as
\begin{equation*}
I(A:C|B_{l}) \leq \varepsilon,
\end{equation*}
with $C := B_{l+1} \backslash B_{l}$.

We can also ask whether Theorem \ref{mthm1} can be extended to higher dimensions, i.e. is any state on a $D$-dim lattice with small $I(A:C|B)$ for $B$ shielding $A$ from $C$ close to thermal? As displayed in Theorem~\ref{thm2}, we may need additional conditions for general graphs. Although we do not know any counter-example, we also could not find any partial result  
whether the additional conditions in those theorems are necessary. 

\section{Acknowledgments}

Part of this work was done when both of us were working in the QuArC group of Microsoft Research. KK thanks Advanced Leading Graduate Course for Photon Science (ALPS) and JSPS KAKENHI Grant Number JP16J05374 for financial support. We thank Matthew Hastings and Michael Kastoryano for useful discussions.

\appendix
\section{Another approach to prove Theorem~\ref{m.mthm2}}
In Sec.~\ref{sec:proof}, we employ a perturbative method to obtain a local operator $O$ determined by the Hamiltonian and a local operator $V$ such that 
\begin{equation*}
e^{-\beta (H+V)}\approx Oe^{-\beta H}O^\dagger\,.
\end{equation*}
The existence of such operator plays a central role in the proof of Theorem~\ref{m.mthm2}. In this appendix, we introduce another approach to obtain similar 
operators, which is based on the previous work by Araki~\cite{Araki1969}. 
The main difference between these approaches is the origins of locality of the operator $O$. In the perturbative approach, the locality is originated in Lieb-Robinson bounds, which restrict real time evolutions of operators. Instead, in Araki\rq{}s approach, locality of $O$ is originated in a restriction on imaginary-time evolutions of $V$. 

Let us consider a 1D spin chain $\Lambda=[-n,n]$, a short-range Hamiltonian $H$ on $\Lambda$, and a local operator $V$. 
We denote the maximum strength of $H$ by $J$. 
A simple algebra show the following relation holds.
\begin{align*}
e^{-\beta (H+V)}&=e^{-\frac{\beta}{2}(H+V)}e^{\frac{\beta}{2}H}e^{-\beta H}e^{\frac{\beta}{2}H}e^{-\frac{\beta}{2}(H+V)}\\
&\equiv E_r(V; H)e^{-\beta H}E_r(V; H)^{\dagger}\,,
\end{align*}
where we denote $e^{-\frac{\beta}{2}(H+V)}e^{\frac{\beta}{2}H}$ by $E_r(V; H)$. 
By denoting $V(\beta)=e^{-\beta H}Ve^{\beta H}$, $E_r(V; H)$ has another form written as~\cite{Araki1969} 
\begin{equation*}
E_r(V; H)=\sum_{n=0}^\infty(-1)^n\int_0^{\frac{\beta}{2}}d\beta_1\int_0^{\beta_1}d\beta_2\cdots\int_0^{\beta_{n-1}}d\beta_nV(\beta_n)\cdots V(\beta_1)\,.
\end{equation*}
Actually, $E_r(V; H)$ can be approximated by a local operator.
\blm~\cite{Araki1969} 
The following statements hold for any region $X\subset [-n,n]$ and any bounded operator $V$ with $supp(V)=[a,b]\subset[-n,n]$.
\begin{itemize}
\item[(i)] There exists a constant $C\geq0$ depending on $\beta$, $J$ and $\|V\|$such that
\begin{equation*}
\|E_r(V; H_X)\| \leq C\,
\end{equation*}
\item[(ii)] There exist constants $C, q\geq0$ depending on $\beta$, $J$ and $\|V\|$ such that
\begin{equation*}
\left\|E_r(V; H_X)-E_r(V;  H_{X\cap[a-l,b+l]})\right\|\leq C\frac{q^{1+\lfloor\frac{l}{2}\rfloor}}{(1+\lfloor\frac{l}{2}\rfloor)!}\,.
\end{equation*}
\end{itemize}
\elm
Since $\log x!\approx x\log x-x$, the denominator grows faster than the numerator with respect to $l$, and thus, the accuracy of the above approximation is exponentially good with respect to $l$. 
Note that similar properties hold for the inverse of $E_r(V; H_X)$, $E_l(V;H_X)\equiv e^{-\frac{\beta}{2}H_X}e^{\frac{\beta}{2}(H_X+V)}$. 
Therefore, by choosing $V=H_{B^M}$, $E_r(V; H)$ and its local approximation play the same role as $O_{ABC}$ and $O_B$ in Sec.~\ref{sec:prth1}.


\begin{thebibliography}{10}

\bibitem{ibinson2008robustness}
Ibinson,~B., Linden,~N. and Winter~A.:
\newblock Robustness of quantum markov chains.
\newblock {\em Comm. Math. Phys.}, 277(2):289--304, 2008.

\bibitem{Fawzi2015}
{Fawzi}~O. and {Renner}~R.:
\newblock Quantum conditional mutual information and approximate markov chains.
\newblock {\em Comm. Math. Phys.}, 340(2):575--611, 2015.

\bibitem{PhysRevLett.100.230501}
Devetak,~I. and Yard~J.:
\newblock Exact cost of redistributing multipartite quantum states.
\newblock {\em Phys. Rev. Lett.}, 100:230501, 2008.

\bibitem{2016arXiv160906994B}
{Berta}, M. and {Brandao}, F.~G.~S.~L. and {Majenz}, C. and {Wilde}, M.~M.:
\newblock {Deconstruction and conditional erasure of quantum correlations}.
\newblock {\em ArXiv e-prints 1609.06994},  2016.

\bibitem{2017arXiv170302903S}
{Sharma},~K., {Wakakuwa}~E., and  {Wilde}, M.~M.:
\newblock {Conditional quantum one-time pad}.
\newblock {\em ArXiv e-prints 1703.02903},  2017.

\bibitem{HCthm71}
{Hammersley},~J.~M. and {Clifford},~P.~E.:
\newblock Markov field on finite graphs and lattices.
\newblock {\em http://www.statslab.cam.ac.uk/~grg/books/hammfest/hamm-cliff.pdf } 1971.

\bibitem{Leifer20081899}
{Leifer}, M.~S. and {Poulin},~D.:
\newblock Quantum graphical models and belief propagation.
\newblock {\em Ann. Phys.}, 323(8):1899 -- 1946, 2008.

\bibitem{2007quant.ph..1029Z}
{Brown}, W. and {Poulin}~D.:
\newblock {Quantum Markov Networks and Commuting Hamiltonians}.
\newblock {\em ArXiv e-prints 1206.0755}, 2012.

\bibitem{2016arXiv160907877B}
{Brandao},~F.~G.~S.~L. and {Kastoryano},~M.~J.:
\newblock {Finite correlation length implies efficient preparation of quantum
  thermal states}.
\newblock {\em ArXiv e-prints1609.07877}, 2016.

\bibitem{PhysRevB.87.155120}
Kim,~I.~H.:
\newblock Determining the structure of the real-space entanglement spectrum
  from approximate conditional independence.
\newblock {\em Phys. Rev. B}, 87:155120, 2013.

\bibitem{Flammia2017limitsstorageof}
Flammia,~S.~T., Haah,~J., Kastoryano,~M.~J. and Kim,~I.~H.:
\newblock Limits on the storage of quantum information in a volume of space.
\newblock {\em {Quantum}}, 1:4, 2017.

\bibitem{PhysRevLett.96.110405}
Levin,~M. and Wen,~X.-G.:
\newblock Detecting topological order in a ground state wave function.
\newblock {\em Phys. Rev. Lett.}, 96:110405, 2006.

\bibitem{upcoming17}
Kato,~K. and {Brandao},~F.~G.~S.~L.:
\newblock Locality of Edge States and Entanglement Spectrum from Strong Subadditivity 
\newblock {\em arXiv preprints}, arXiv:1804.05457, 2018.

\bibitem{PhysRevLett.100.070502}
Wolf,~M.~M., Verstraete,~F., Hastings,~M.~B. and Cirac,~J.~I.:
\newblock Area laws in quantum systems: Mutual information and correlations.
\newblock {\em Phys. Rev. Lett.}, 100:070502, 2008.

\bibitem{hastings2006solving}
Hastings,~M.~B.:
\newblock Solving gapped hamiltonians locally.
\newblock {\em Phys. Rev. B}, 73(8):085115, 2006.

\bibitem{PhysRevB.94.155125}
Swingle, B. and McGreevy, J.:
\newblock Mixed $s$-sourcery: Building many-body states using bubbles of nothing. 
\newblock{\em Phys. Rev. B}, 94(15):155125, 2016.

\bibitem{KastoryanoBrandao16}
Branda o, F. G. S. L.:
\newblock Finite correlation length implies efficient preparation of quantum thermal states
\newblock {\em arXiv prepromts}, arXiv:1609.07877, 2016.

\bibitem{PhysRev.106.620}
{Jaynes},~E.~T.:
\newblock Information theory and statistical mechanics.
\newblock {\em Phys. Rev.}, 106:620--630, 1957.

\bibitem{PhysRev.108.171}
{Jaynes},~E.~T.:
\newblock Information theory and statistical mechanics. ii.
\newblock {\em Phys. Rev.}, 108:171--190, 1957.

\bibitem{2010.5671W}
{Weis},~S.:
\newblock {Information topologies on non-commutative state spaces}.
\newblock {\em J. Conv. Anal.}, 21(2):339--399, 2014.

\bibitem{weismaxent2015}
{Weis},~S.:
\newblock The maxent extension of a quantum gibbs family, convex geometry and
  geodesics.
\newblock {\em AIP Conference Proceedings}, 1641, 2015.

\bibitem{AF04}
Alicki~R. and Fannes~M.:
\newblock Continuity of quantum conditional information
\newblock {\em J. Phys. A: Math. Gen.}, 37 (2004) L55

\bibitem{PhysRevB.76.201102}
Hastings,~M.~B.:
\newblock Quantum belief propagation: An algorithm for thermal quantum systems.
\newblock {\em Phys. Rev. B}, 76:201102, 2007.

\bibitem{PhysRevB.86.245116}
Kim,~I.~H.:
\newblock Perturbative analysis of topological entanglement entropy from
  conditional independence.
\newblock {\em Phys. Rev. B}, 86:245116, Dec 2012.

\bibitem{Araki1969}
Araki,~H.:
\newblock Gibbs states of a one dimensional quantum lattice.
\newblock {\em Comm. Math. Phys.}, 14(2):120--157, 1969.

\bibitem{lieb1972}
Lieb,~E.~H.  and Robinson,~D.~W.:
\newblock The finite group velocity of quantum spin systems.
\newblock {\em Comm. Math. Phys.}, 28(3):251--257, 1972.

\bibitem{Araki1973}
Araki,~H.:
\newblock Expansional in banach algebras.
\newblock {\em Ann. sci. Ecole Norm. S.}, 6(1):67--84, 1973.

\bibitem{Brandao2015}
Brandao,~F.~G. S.~L. and Horodecki,~M.:
\newblock Exponential decay of correlations implies area law.
\newblock {\em Comm. Math. Phys.}, 333(2):761--798, 2015.

\end{thebibliography}
\end{document}